\documentclass{nature_max}

\usepackage{graphicx}
\usepackage{amsmath}
\usepackage{bm}
\usepackage{color}
\title{Evidence of ideal excitonic insulator in 
bulk MoS$_2$ under pressure }

\author{S. Samaneh Ataei$^1$, Daniele Varsano$^1$, 
        Elisa Molinari$^{1,2}$
        \& Massimo Rontani$^1$}

\begin{document}

\maketitle

\begin{affiliations}
 \item CNR-NANO, Via Campi 213a, 41125 Modena, Italy.
 \item Dipartimento di Scienze Fisiche, Informatiche e Matematiche (FIM), 
       Universit{\`a} degli Studi di Modena e Reggio Emilia, 41125 Modena, 
       Italy.
\end{affiliations}

\begin{abstract}
Spontaneous condensation of excitons is a long sought phenomenon analogous to the condensation of Cooper pairs in a superconductor. It is expected to occur in a semiconductor at thermodynamic equilibrium if the binding energy of the excitons---electron ($e$) and hole ($h$) pairs interacting by Coulomb force---overcomes the band gap, giving rise to a new phase: the ‘excitonic insulator’ (EI). Transition metal dichalcogenides are excellent candidates for the EI realization because of reduced Coulomb screening, and indeed a structural phase transition was observed in few-layer systems. However, previous work could not disentangle to which extent the origin of the transition was in the formation of bound excitons or in the softening of a phonon.
Here we focus on bulk MoS$_2$ and demonstrate theoretically that at high pressure it is prone to the condensation of genuine excitons of finite momentum, whereas the phonon dispersion remains regular. Starting from first-principles many-body perturbation theory, we also predict that the self-consistent electronic charge density of the EI sustains an out-of-plane permanent electric dipole moment with an antiferroelectric texture in the layer plane: At the onset of the EI phase, those optical phonons that share the exciton momentum provide a unique Raman fingerprint for the EI formation.
Finally, we identify such fingerprint in a Raman feature that was previously observed experimentally, thus providing direct spectroscopic confirmation of an ideal excitonic insulator phase in bulk MoS$_2$ above 30 GPa.
\end{abstract}

\newpage

The long-sought excitonic insulator (EI) is a permanent Bose-Einstein condensate of excitons in the absence of optical excitation, hosted in a narrow-gap semiconductor or a semimetal   \cite{Keldysh1964,Cloizeaux1965,Kohn1967,Kohn1967b}. As the exciton condensate shares similarities with the superconductor ground state  \cite{BCS1957}, it may exhibit macroscopic quantum coherence and exotic low-energy excitations \cite{Halperin1968,Guseinov1973,Portengen1996b,Stringari2003,Eisenstein2004,Littlewood2008,Rontani2013}. These intriguing features are linked to the arbitrariness of the phase of the condensate wave function, $\varphi$ (defined in Eq.~2 below): whereas in the superconductor this phase degeneracy is protected by the conservation of electronic charge, in the EI it is contingent on the preservation of excitons \cite{Guseinov1973,Littlewood2008}, and hence lifted by those terms in the Hamiltonian that annihilate or create $e$-$h$ pairs. This is the case of $e$-phonon \cite{Volkov1973} and spin-orbit \cite{Varsano2020} interactions, which pin $\varphi$ while hybridizing conduction and valence bands [remarkably, spin-orbit coupling provides excitonic insulators with topological properties \cite{Varsano2020}]. So far, the most accomplished EIs were realized in bilayer heterostructures in the presence of a magnetic field, requiring both low temperature and complex engineering to maximize the impact of $e$-$h$ correlations  as well as the degeneracy of $\varphi$ \cite{Eisenstein2004,Nandi2012}. A related concept aims to achieve the temporary condensation of indirect excitons, made of spatially separated $e$ and $h$, through the optical pumping of artificial bilayers designed to maximize the exciton lifetime \cite{Butov2002a,High2012,Anankine2017}.   

Recently, layered 
materials \cite{Rohwer2011,Kogar-Abbamonte_2017,Kono2017,Kaiser2018} renewed the promise of the EI because of the enhanced Coulomb interactions, and hence exciton binding, due to their reduced dimensionality. In particular, the indirect character of excitons---in reciprocal \cite{Rohwer2011,Kogar-Abbamonte_2017} and real \cite{Kono2017,Kaiser2018} space for TiSe$_2$ and Ta$_2$NiSe$_5$, respectively---prevented $e$-$h$ pairs from dissociation due to screening. In those systems, the putative transition to the EI was accompanied by a lattice instability  \cite{DiSalvo1976,Hedayat2019,Zhou2019,Nakano2018,Yan2019} when lowering the temperature---a singularity in the phonon density of states at vanishing energy---that in turn created $e$-$h$ pairs through $e$-phonon interaction. In contrast, the transition to the ideal EI is purely electronic, with only small adjustments of the lattice  \cite{Kohn1967b,Kohn1970}.

\begin{figure*}[htbp]
\centering
%\setlength{\unitlength}{1 cm}
%\begin{picture}(16,10)
\includegraphics[width=11.4cm]{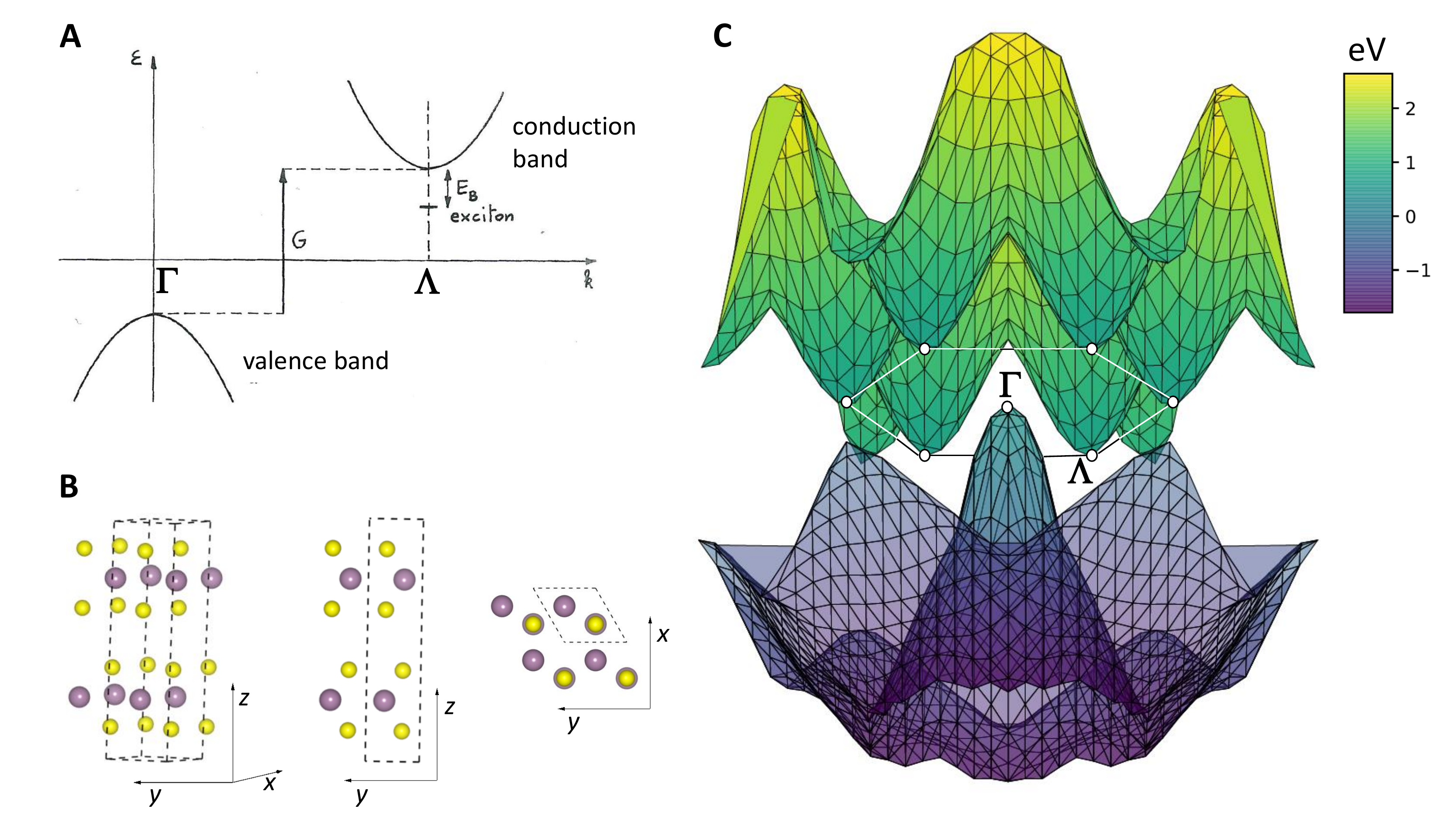}
%\put(0.0,0.5){\includegraphics[trim=0cm 0cm 0 0,clip=true,width=17.8cm]{./Fig1.pdf}}
%\end{picture}
\caption{
{\bf Indirect-gap MoS$_2$ as a candidate excitonic insulator.}
({\bf A}) Sketch of the excitonic insulator instability, adapted from Walter Kohn's original proposal, Ref.~\cite{Kohn1967b}. An exciton binds an electron at the conduction band bottom, located at $\Lambda$ in {\bf k} space, with a hole at the valence band top at $\Gamma$. If the exciton binding energy, $E_b$, is larger than the indirect gap, $G$, then the system is unstable against the spontaneous generation of excitons. The reconstructed many-body ground state---a condensate of excitons at thermodynamic equilibrium---is the excitonic insulator. 
({\bf B}) Model of the 2$H_c$ crystal structure from different views. The violet (yellow) colour labels Mo (S) atoms. The dashed frame appearing in side and top views is the unitary cell of the  layered structure, with $a$ and $c$ being the in- and out-of-plane lattice constants, respectively.  
({\bf C}) Lowest conduction and topmost valence 
energy band as a function of wave vector in the $k_z=0$ plane, as obtained from first-principles many-body perturbation theory (GW) at a pressure of 34 GPa.
}
\end{figure*}

Here, we follow an early suggestion by
Hromadov\'a {\it et al.}~\cite{tosatti} and focus on bulk MoS$_2$ under hydrostatic pressure  \cite{tosatti,Chi2014,Nayak2014,Chi2018}. We use many-body perturbation theory from first principles  \cite{Onida-Reining-Rubio_2002,Varsano-Rontani2017} to demonstrate that MoS$_2$ is 
unstable against exciton condensation but stable against lattice distortion. Bulding a self-consistent effective-mass model on top of ab initio calculations, we show that the true ground state is an ideal, anti-ferrolectric EI with a distinctive Raman fingerprint that has already been observed \cite{Cao2018}.    

In bulk MoS$_2$, the pressure ($P$) closes the indirect gap, $G$, between the top of the filled valence band---located at the center of the Brillouin zone ($\Gamma$ point), and the bottom of the six-degenerate valleys of the empty conduction band---placed at $\Lambda$ points (approximately midway between $\Gamma$ and K, see Fig.~1C for $P=$ 34 GPa). The energy landscape along one of the $\Gamma\Lambda$ cuts (sketched in Fig.~1A) favours the Coulomb binding of an $e$, located at $\Lambda$, with a $h$, placed at $\Gamma$, creating an exciton of finite momentum $|${\bf q}$|$ $=\Gamma\Lambda$ and binding energy $E_b$. Whereas ordinarily $E_b<G$, it may occur that $E_b>G$ above a critical pressure, a condition that makes the semiconductor unstable against the condensation of excitons. This is actually the case, as we show below from first principles.

So far, ultra high pressure has been used as a handle to make MoS$_2$ superconducting \cite{Chi2018} (at $P\sim$ 90 GPa), though the pairing mechanism remains unclear  \cite{Ge2013,Roldan2013,Rosner2014}. The putative EI must be searched at lower pressure ($P\sim$ 25 GPa), close to the semiconductor-semimetal transition that was observed by several groups  \cite{aksoy,Chi2014,Nayak2014,Bandaru2014,Zhuang2017}.
Near this boundary, theory  \cite{tosatti,ordejon}---including our own calculations (SI Appendix, Fig.~S1)---predicts an isostructural transition from the 2$H_c$ (Fig.~1B) to the 2$H_a$ (SI Appendix, Fig.~S2) phase, which does not affect the crystal space group $D^4_{6h}$, as the two structures transform into each other through the sliding of the layers in the unit cell (the layer unit is made of one Mo and two S atoms, represented respectively by violet and yellow balls in the sketch of Fig.~1B). Raman and x-ray spectroscopic observations \cite{Chi2014,Bandaru2014,Chi2018,Cao2018,Goncharov2020} suggest that 2$H_c$ and 2$H_a$ phases coexist in diamond-anvil cells, in a range that varies between 25 and 50 GPa in powders but has narrower extension ($\sim$ 4 GPa) in single crystals. Importantly, we find that both 2$H_c$ and 2$H_a$ polytypes experience a similar excitonic instability---unrelated to the structural transition, as the electronic bands of the two phases are basically identical close to the Fermi energy. Below, we discuss the 2$H_c$ stacking and leave the analysis of 2$H_a$ to the SI Appendix, Figs.~S3 and S4. 

\begin{figure*}[htbp]
\centering
%\setlength{\unitlength}{1 cm}
%\begin{picture}(16,9.5)
%\put(0.2,-0.5){\includegraphics[trim=2.5cm 0 2cm 0cm,clip=true,width=7.5cm]{./N_01_HD.jpg}}
%\put(9.1,0.2){\includegraphics[trim=5cm 1.5cm 0 0,clip=true,width=7.5cm]{./Fig1b_new.pdf}}
%\includegraphics[trim=2.5cm 0 2cm 0cm,clip=true,width=5.5cm]{./N_01_HD.jpg}
%\qquad\qquad
%\put(0.0,0.2){\includegraphics[trim=0cm 0cm 0 0,clip=true,width=16.0cm]{./Fig2.pdf}}
\includegraphics[width=11.4cm]{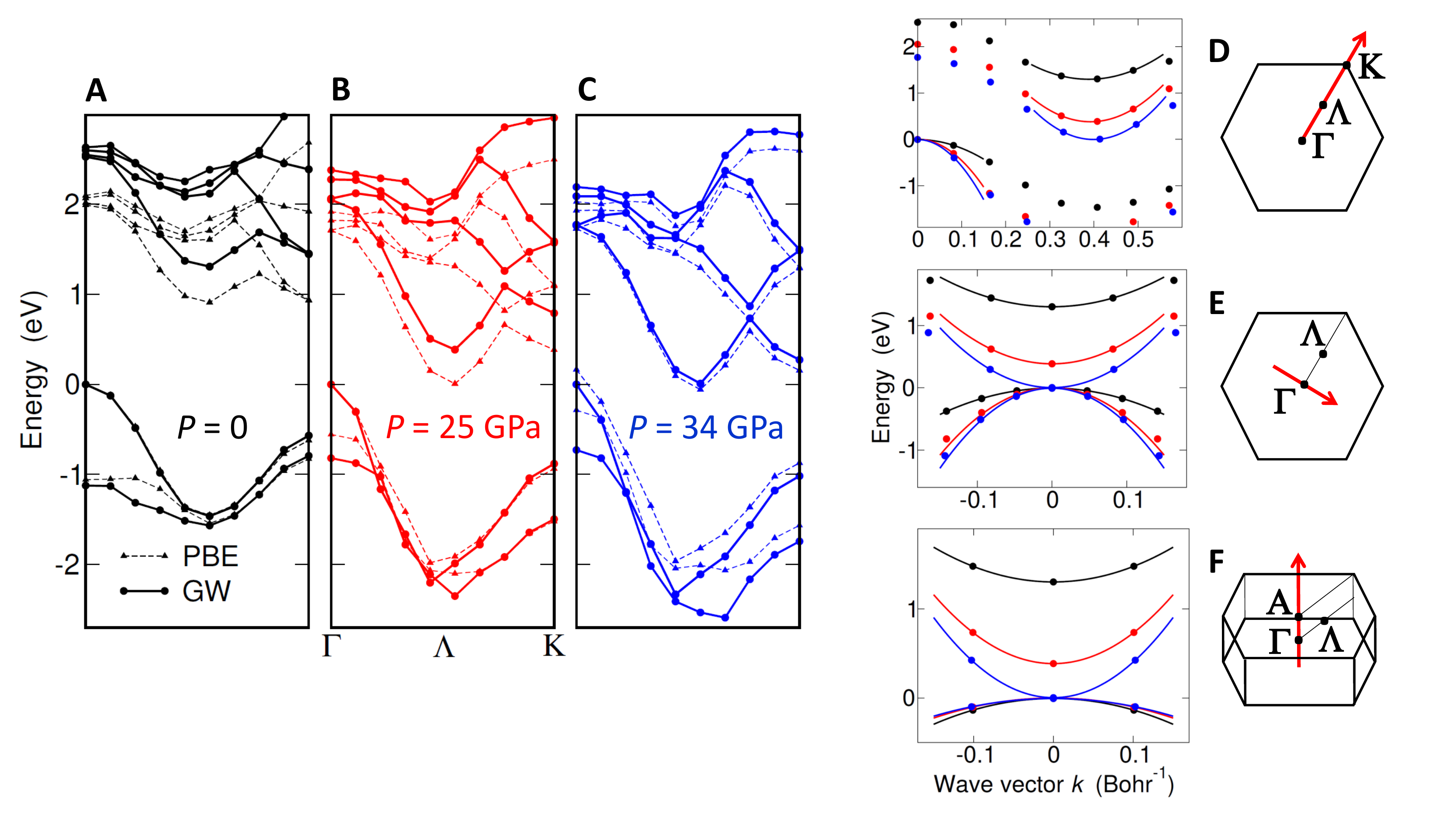}
%\put(0.5,7.5){\bf{a}}
%\put(0.5,3.0){\bf{b}}
%\put(4.2,7.5){\bf{c}}
%\put(4.2,3.0){\bf{d}}
%\put(10.2,5.9){\bf{e}}
%\end{picture}
\caption{
{\bf Closing the gap by applying pressure. }
({\bf A} to {\bf C}) Band structure
along the $\Gamma-\Lambda-$ K cut of the Brillouin zone at pressure $P=$ 0 (panel A), 25 (panel B), 34 GPa (panel C). Band energies obtained from first principles including the quasiparticle GW corrections beyond DFT (circles) are compared to bare DFT data (triangles, PBE functional). Lines are guides to the eye. 
({\bf D} to {\bf F}) Band dispersion of conduction and valence bands close to $\Lambda$ and $\Gamma$ points, respectively, for $P=$ 0 GPa (black colour), 25 GPa (red), and 34 GPa (blue). In panels E and F the conduction band has been rigidly translated by the wave vector $-\Vec{\Gamma\Lambda}$. GW predictions (dots) are shown together with effective-mass fits (curves). The directions of the cuts (shown as red arrows in the Brillouin zone) are the principal axes of the effective-mass tensor, two being in the $k_z=0$ plane (panels D and E) and one being parallel to the $k_z$ axis (panel F).  
}
\end{figure*}

\section*{Results}

The indirect gap of 2$H_c$--MoS$_2$ is sensitive to pressure, as its value drops from 1.31 eV at $P=0$ (Fig.~2A) to only 9 meV at $P = 34$ GPa (Fig.~2C), close to the semimetal limit. With respect to the accurate band structure calculated within the GW approximation (circles in Fig.~2A to C, see Methods), density functional theory (triangles) underestimates the gap of about 0.4 eV at $P=0$. 
However, as pressure reduces the out-of-plane lattice parameter $c$ (SI Appendix, Fig.~S1), forcing sulfur orbitals belonging to adjacent layers to overlap \cite{guo}, virtual $e$-$h$ pairs start tunnelling among layers, screening effectively Coulomb interaction at long wavelength. This reduces the GW energy correction to DFT bandgap, as evident in Fig.~2C.
Consistently, the conduction band increases its dispersion along the $k_z$ direction (Fig.~2F), as well as the other
axes of the effective mass tensor (Figs.~2D to E; dots and lines are GW data and effective-mass fits, respectively).
Overall, the semiconductor becomes progressively more isotropic as it turns into a semimetal, loosing its two-dimensional character.   

\begin{figure*}[htbp]
\centering
%\setlength{\unitlength}{1 cm}
%\begin{picture}(16,9)
%\put(0.2,-0.5){\includegraphics[trim=2.5cm 0 2cm 0cm,clip=true,width=7.5cm]{./N_01_HD.jpg}}
%\put(9.1,0.2){\includegraphics[trim=5cm 1.5cm 0 0,clip=true,width=7.5cm]{./Fig1b_new.pdf}}
%\includegraphics[trim=2.5cm 0 2cm 0cm,clip=true,width=5.5cm]{./N_01_HD.jpg}
%\qquad\qquad
\includegraphics[trim=0cm 0cm 0 0,clip=true,width=14.8cm]{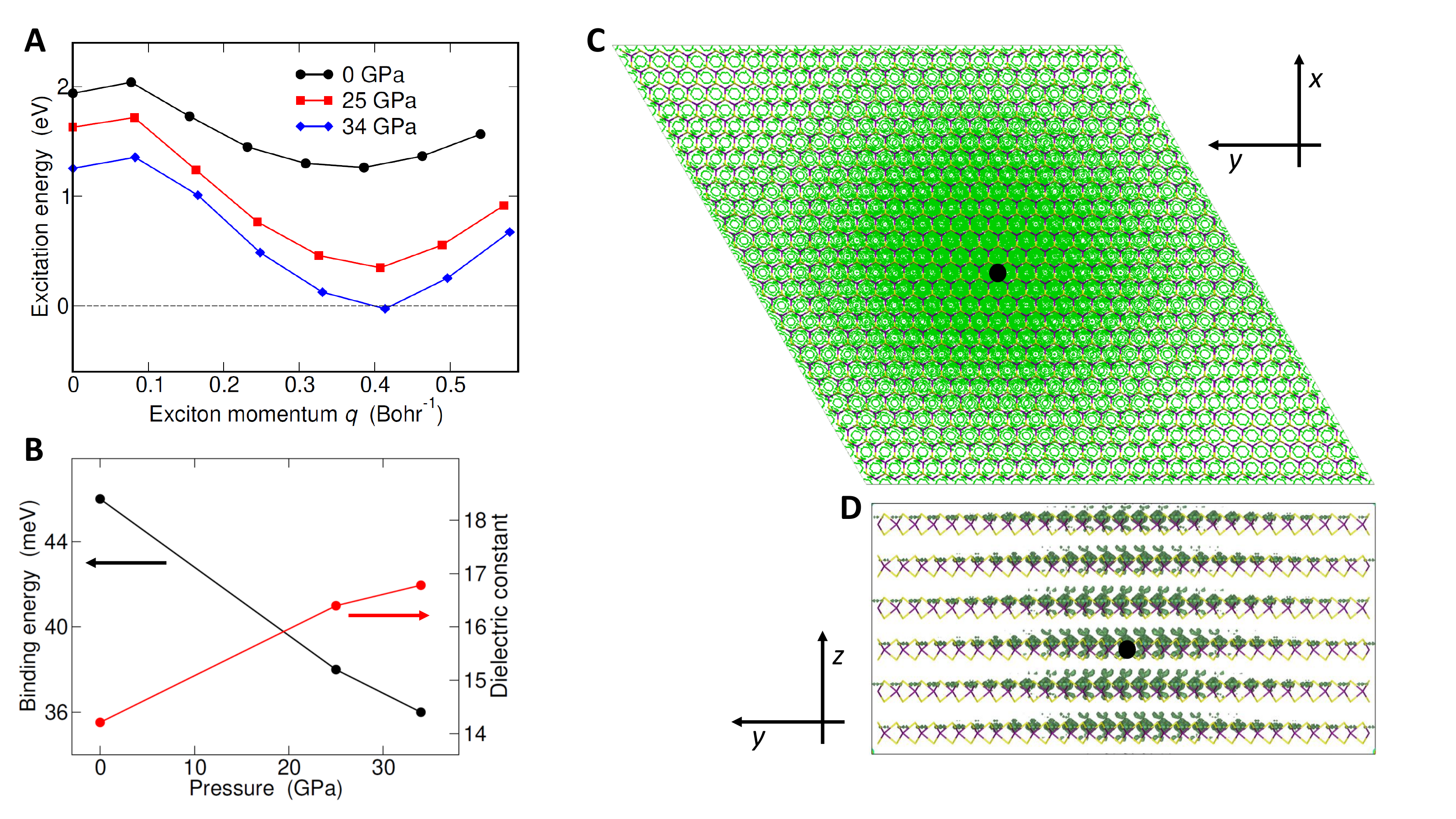}
%\includegraphics[width=11.4cm]{./Fig3.pdf}
%\put(0.5,7.5){\bf{a}}
%\put(0.5,3.0){\bf{b}}
%\put(4.2,7.5){\bf{c}}
%\put(4.2,3.0){\bf{d}}
%\put(10.2,5.9){\bf{e}}
%\end{picture}
\caption{
{\bf Excitonic instability. }
({\bf A})  Excitation energy of the lowest exciton vs center-of-mass momentum {\bf q} along the $\Gamma$K direction. Data are obtained from first principles for $P=$ 0 (dots), 25 (squares), 34 GPa (diamonds). Note that the K point position expressed in units of Bohr$^{-1}$ 
shifts with $P$. 
Solid lines are guides to the eye. 
At $P=$ 34 GPa the excitation energy is negative for $q=\Gamma\Lambda$, which points to the instability against exciton condensation (the dashed line highlights the energy zero). 
({\bf B}) Binding energy of the exciton having momentum $q=\Gamma\Lambda$ (black circles, left vertical axis) and macroscopic static dielectric constant (red circles, right axis) vs $P$. The latter is obtained through the inverse dielectric matrix, as $1/\,[\epsilon^{-1}(\mathbf{q}=0)]_{\mathbf{G}=\mathbf{G'}=0}$ ({\bf G} is the reciprocal lattice vector). ({\bf C} and {\bf D}) Wave function square modulus of the lowest exciton with $q=\Gamma\Lambda$ at $P=0$. The plot shows the conditional probability to locate the bound electron (green contour map), provided the hole position is fixed (black dot), either in (panel C) or out (panel D) of plane. The violet (yellow) colour in the stick-and-ball skeleton points to Mo (S) atoms.  
}
\end{figure*}

\section*{Exciton binding and instability} 

The exciton candidate for the instability has a finite center-of-mass momentum {\bf q}, i.e., it travels in space. We compute its excitation energy---the difference between the GW bandgap and the binding energy---by solving the Bethe-Salpeter equation from first principles (Methods). The dispersion exhibits a dip for $q=\Lambda$, whose energy is first positive at $P=0$ (1.26 eV, black dots in Fig.~3A) but then quickly lowers with $P$, eventually changing sign close to the semimetal threshold ($-27$ meV at $P=34$ GPa, blue dots). This negative value signals that excitons spontaneously form, which leads to a reconstructed many-body phase of lower energy.

The softening of the exciton shown in Fig.~3A validates from first principles the seminal prediction by des Cloizeaux \cite{Cloizeaux1965}: the binding energy remains finite even if the gap vanishes, as explicitly shown in Fig.~3B (black dots). The reason is that conduction and valence band profiles are almost unaffected by $P$ (Fig.~2),  
as the band edges are displaced in {\bf k} space, which prevents the macroscopic dielectric constant from diverging (red dots in Fig.~3B).
Were the closing gap direct, metal-like screening would dissociate the exciton.

The square modulus of the exciton wave function is illustrated in Figs.~3C and D, as the conditional probability density to locate the bound electron (green contour map), provided the hole is fixed (black dot). Note that the center-of-mass motion does not appear in this frame. The probability extends tens of Angstroms---the feature of Wannier excitons familiar from bulk semiconductors---both in- and out-of-plane, as apparent in panels C and D, respectively (the Bohr radius is 50 {\AA} at 34 GPa, as shown in SI Appendix, Fig.~S5). The exciton becomes lighter and more isotropic with pressure, i.e., more delocalized in real space (here shown at $P=0$).

\section*{Two-band model} 

The major source of numerical error is the finite sampling of the Brillouin zone \cite{Varsano2020}, since the exciton is significantly localized in {\bf k} space while the computational load prevents us from refining the mesh (Methods). However, the specific features of the exciton provide us with a workaround, since: (i) the wave function is spanned essentially by those $e$ and $h$ states that are close to the edges of the lowest conduction and highest valence band, respectively (Fig.~1A); (ii) the spin degree of freedom is irrelevant, the exciton energy being four-fold degenerate within numerical accuracy (spin-orbit coupling is fully included in the calculation). Therefore, we may afford ultradense {\bf k}-space sampling by replacing the first-principles Bethe-Salpeter equation with its spinless two-band counterpart within the effective mass approximation  \cite{Varsano-Rontani2017}, the mass tensor being extracted from Figs.~2D to F and the dielectric constant from Fig.~3B (Methods and SI Appendix, Fig.~S5). The resulting excitation energy, at the semimetal threshold, is $\approx -8$ meV.             

\begin{figure*}[htbp]
\centering
%\setlength{\unitlength}{1 cm}
%\begin{picture}(16,9)
%\put(0.2,-0.5){\includegraphics[trim=2.5cm 0 2cm 0cm,clip=true,width=7.5cm]{./N_01_HD.jpg}}
%\put(9.1,0.2){\includegraphics[trim=5cm 1.5cm 0 0,clip=true,width=7.5cm]{./Fig1b_new.pdf}}
%\includegraphics[trim=2.5cm 0 2cm 0cm,clip=true,width=5.5cm]{./N_01_HD.jpg}
%\qquad\qquad
%\put(0.0,0.2){\includegraphics[trim=0cm 0cm 0 0,clip=true,width=16.0cm]{./Fig4.pdf}}
\includegraphics[trim=0cm 0cm 0cm 0cm,clip=true,width=14.8cm]{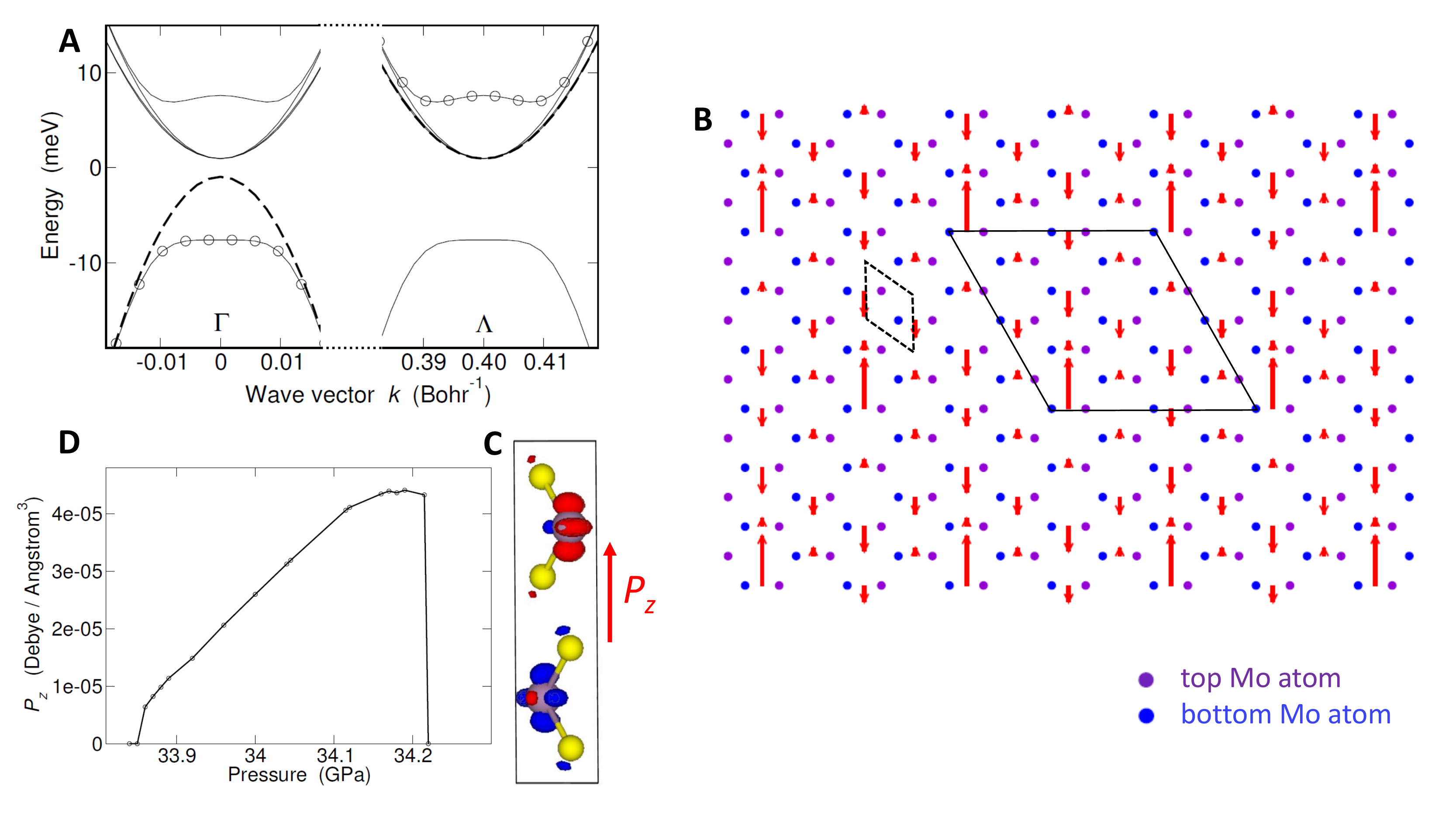}
%\includegraphics[trim=0cm 0cm 0 0,clip=true,width=11.4cm]{./Fig4.pdf}
%\put(0.5,7.5){\bf{a}}
%\put(0.5,3.0){\bf{b}}
%\put(4.2,7.5){\bf{c}}
%\put(4.2,3.0){\bf{d}}
%\put(10.2,5.9){\bf{e}}
%\end{picture}
\caption{
{\bf Anti-ferroelectric excitonic insulator.} \tiny
({\bf A}) Band structure of the excitonic (solid curves) and pristine (dashed curves) insulator along one of the six equivalent $\Gamma-\Lambda$ directions in the $k_z=0$ plane of the Brillouin zone at $P=$ 34 GPa. The original conduction bands are folded from $\Lambda$ valleys to $\Gamma$, and renormalized together with the valence band. The new band structure at $\Gamma$ is replicated at $\Lambda$, since both $\Gamma$ and $\Lambda$ points belong to the EI reciprocal lattice. Apart from spin degeneracy, EI renormalized bands exhibit an additional orbital degeneracy reminescent of the pristine multivalley structure:
bands (solid curves) from top to bottom are respectively one-, three-, two-, and one-fold degenerate, respectively. The two-fold degenerate band, which overlaps with the pristine conduction band (dashed curve), is actually split due to the tiny anisotropy of $\Delta_{\text{\bf k}}$ in the $k_x$, $k_y$ plane (splitting hardly visible in the plot). Circles point to EI conduction and valence bands obtained within the two-band model.  ({\bf B}) Anti-ferroelectric structure. A permanent out-of-plane electric dipole, $P_z(x,y)$, spontaneously develops and exhibits an in-plane modulation that breaks inversion symmetry. This dipole, which averages to zero over the unitary cell of the superstructure (solid frame),  is perpendicular to the plane and depicted as a red arrow of varying sign and modulus in the figure. The superstructure cell contains 72 atoms against 6 of the original cell (dashed frame) [here $\Gamma\Lambda\approx 2\pi/(3a)$]. Violet and blue dots are Mo atoms respectively in the top and bottom layer (S atoms are not shown). ({\bf C}) Overlap charge density of the periodic part of pristine conduction and valence band Bloch states, respectively at $\Lambda$ and $\Gamma$, shown in the 2$H_c$ cell. The red (blue) color points to a surplus (deficit) of charge. The depicted charge displacement, which is associated with the polarization of condensed excitons, is the origin of the permanent dipole $P_z$ shown in panel B. ({\bf D}) Maximum local value of $P_z$ vs $P$.}
\end{figure*}

\section*{The excitonic insulator phase} 

Close to the semiconductor-semimetal boundary, the ground state undergoes a reconstruction from the `normal' phase, $\left|\Phi_0\right>$, which is either insulating or semimetallic, to the excitonic insulator, $\left|\Psi_{\text{EI}}\right>$. 
In the following, we highlight the essential features of $\left|\Psi_{\text{EI}}\right>$ within the simpler two-band model (as a mnemonic, we adopt the apex `0' to identify quantities of interest defined within this model). Then, we take into account the EI multivalley nature by adapting the theory first proposed for the candidate material TiSe$_2$ \cite{Monney2009}. 

Within the two-band model \cite{Kohn1967},  $\left|\Psi_{\text{EI}}^0\right>$ is formally analogous to the superconductor wave function \cite{BCS1957}, 
\begin{equation}
\left|\Psi_{\text{EI}}^0\right> = \prod_{\text{\bf k}} [u_{\text{\bf k}}^0 + v_{\text{\bf k}}^0e^{-i\varphi}\,\hat{b}^+_{\text{\bf k}}\hat{a}_{\text{\bf k}} ]\left|\Phi_0\right>,   
\end{equation}
provided the Cooper pairs of the metal are replaced with the $e$-$h$ pair excitations of the normal state, $\hat{b}^+_{\text{\bf k}}\hat{a}_{\text{\bf k}}\left|\Phi_0\right>$. Here $\hat{b}^+_{\text{\bf k}}$ creates an electron with momentum {\bf k} $+\,\Vec{\Gamma\Lambda}$ and energy $\varepsilon_b(\text{\bf k})$ in the conduction band, 
$\hat{a}_{\text{\bf k}}$ annihilates an electron with momentum {\bf k}
and energy $\varepsilon_a(\text{\bf k})$ in the valence band,
$u_{\text{\bf k}}^0$ and $v_{\text{\bf k}}^0$ are positive coherence factors [$(u^0_{\text{\bf k}})^2 + (v^0_{\text{\bf k}})^2=1$], and
$\varphi$ is the phase of of the condensate wave function, $\zeta^0_{\text{\bf k}}=u_{\text{\bf k}}^0v_{\text{\bf k}}^0e^{i\varphi}=\Delta_{\text{\bf k}}^0 / 2E_{\text{\bf k}}$, with $\Delta_{\text{\bf k}}^0$ being the excitonic gap function and $E_{\text{\bf k}} =
\{[\varepsilon_b(\text{\bf k})-\varepsilon_a(\text{\bf k})]^2/4 + | \Delta_{\text{\bf k}}^0|^2\}^{1/2}$. The value of $\varphi$ is---ideally---arbitrary and solely fixed by the spontaneous breaking of the conservation law for $e$-$h$ pairs, as
\begin{equation}
\left<\Psi_{\text{EI}}^0\right| \hat{b}^+_{\text{\bf k}}\hat{a}_{\text{\bf k}} \left|\Psi_{\text{EI}}^0\right> = \left|\zeta^0_{\text{\bf k}}\right|e^{i\varphi}.     
\end{equation}
The EI band structure is obtained by solving the pseudo Bethe-Salpeter equation for $\zeta^0_{\text{\bf k}}$ self-consistently, 
\begin{equation}
2E_{\text{\bf k}}\,\zeta^0_{\text{\bf k}}-\sum_{\text{\bf k}'}
W(\text{\bf k}-\text{\bf k}')\,\zeta^0_{\text{\bf k}'}\;=\;0,
\end{equation} 
where $W({\text{\bf q}})$ is the screened Coulomb interaction and the minimum value of $2E_{\text{\bf k}}$ is the bandgap (Methods). Reassuringly, Eq.~3
turns into the Bethe-Salpeter equation for the zero-energy exciton at the onset of the EI phase ($\Delta^0_{\text{\bf k}}\rightarrow 0+$).
As a consequence of the condensation energy gain, the EI conduction and valence bands (circles in Fig.~4A) are flattened and distorted with respect to those of the pristine semiconductor (dashed curves), the gap widening by $\approx 15$ meV at $P=34$ GPa. 

\section*{Multivalley effects}

As $e$-$h$ pairs may be formed by exciting an electron from the valence band to any one of the six conduction band valleys, $\Lambda_i$, the condensate wave function is multi-component \cite{Monney2009}, $\left<\Psi_{\text{EI}}\right| \hat{b}^+_{i\text{\bf k}}\hat{a}_{\text{\bf k}} \left|\Psi_{\text{EI}}\right>=\zeta_{i\text{\bf k}}$, with $\hat{b}^+_{i\text{\bf k}}$ creating an electron with momentum {\bf k} $+\,\Vec{\Gamma\Lambda}_i$ and energy $\varepsilon_{ib}(\text{\bf k})$ in the $i$th valley ($i=1,\ldots,6$). In principle, one must solve up to six coupled  equations for $\zeta_{i\text{\bf k}}$ to account for the distortion of the condensate in {\bf k} space, due to intervalley coupling. Nevertheless, we note that $\Delta_{\text{\bf k}}^0$ has hardly any angular dependence in the $k_x$, $k_y$ plane (the maximum amplitude of the azimuthal modulation is smaller than 0.07 meV, see SI Appendix, Fig.~S6), whereas $\varepsilon_{ib}(\text{\bf k})$ depends on the angle between $\Vec{\Gamma\Lambda}_i$ and ($k_x$,$k_y$) due to mass anisotropy. As Coulomb interaction protects the cylindrical symmetry of $\zeta_{i\text{\bf k}}$, and since the bare-band anisotropy has negligible effect at valley bottom $\text{\bf k}\approx 0$ (where the value of $\zeta_{i\text{\bf k}}$ is largest), we neglect the
azimuthal dependence of $\zeta_{i\text{\bf k}}$ and obtain (Methods):
\begin{equation}
\zeta_{i\text{\bf k}} = \left<\Psi_{\text{EI}}\right| 
\hat{b}^+_{i\text{\bf k}}\hat{a}_{\text{\bf k}} \left|\Psi_{\text{EI}}\right> = \frac{1}{\sqrt{6}}\,u_{\text{\bf k}}^0v_{\text{\bf k}}^0\, e^{i{\varphi}_i},
\quad i=1,\ldots,6.
\end{equation}
Here only the magnitude of $\zeta_{i\text{\bf k}}$ is fixed (from the self-consistent solution of equation 3), whereas the six phases $\varphi_i$ remain undetermined. This is sufficient to compute the band structure of the EI (Fig.~4A), as the ground state energy is independent from $\varphi_i$ (Methods).

There are now one valence and six conduction bands (solid thin lines in Fig.~4A), in place of the two bands (circles) of the superconductor-like model. Some of the conduction bands are degenerate, the degeneracy being respectively one, three, two, and one, from the topmost conduction to the valence band. Importantly, the band structure at $\Gamma$ is replicated at $\Lambda$, as the electronic charge exhibits a super-modulation in real space that we discuss below, the corresponding unit cell (solid frame in Fig.~4B) being larger than the cell of the crystal lattice (dashed frame). As a consequence, bands are folded into the smaller Brillouin zone (SI Appendix, Fig.~S6), changing the gap from indirect to direct.     
Only the valence and topmost conduction bands repel each other, in agreement with the two-band model (circles), whereas the remaining bands,
which are unaffected by the presence of the exciton condensate,
replicate at $\Gamma$ the bare valleys and hence reduce the direct gap.
Since the location of the valence band top is slightly displaced from $\Gamma$ along the $k_z$ axis (SI Appendix, Fig.~S7), by $\sim 0.2$ Bohr$^{-1}$, the actual EI gap is indirect and around $\sim$ 5 meV, smaller than the direct gap at $\Gamma$. Note that in Fig.~4A the two-fold degenerate band, which almost overlaps with the bare conduction band (dashed curve), splits due to the tiny anisotropy of $\Delta_{\text{\bf k}}$ in the $k_x$, $k_y$ plane (the splitting is hardly visible in the plot).

\section*{Anti-ferroelectric excitonic insulator}
 
The EI ground state is invariant under time reversal, hence the phases of the condensate components that live in two antipodal valleys must have opposite sign  (modulus a multiple integer of $2\pi$), i.e., $\varphi_1 = - \varphi_4$, $\varphi_3=-\varphi_6$, and $\varphi_5=-\varphi_2$ (see SI Appendix, Fig.~S6 and Methods). This constraint leads to the formation of a purely electronic, self-sustained charge density wave, $\Delta\varrho(\text{\bf r})$,
which breaks the inversion symmetry of the pristine crystal (the proof is given in the Methods). The total wave $\Delta\varrho$ is the coherent superposition of three contributions,
$\Delta\varrho=\Delta\varrho_{1,4}+\Delta\varrho_{3,6}+\Delta\varrho_{5,2}$, each one originating from a couple of antipodal valleys. For example,
\begin{eqnarray}
\Delta\varrho_{1,4}(\text{\bf r}) & = & \frac{8}{\sqrt{6}}
\left[\sum_{\text{\bf k}}\,u_{\text{\bf k}}^0v_{\text{\bf k}}^0\right]\quad\times \nonumber \\
&& \text{Re}\!\left\{\psi_{\Gamma}(\text{\bf r})\,\psi^*_{\Lambda_1}\!(\text{\bf r})\,
\exp{\![-i(\Vec{\Gamma\Lambda}_1 \!\cdot \!\text{\bf r} - \varphi_1 )]}\right\} 
,
\end{eqnarray}
exhibits the new periodicity $2\pi/|\Vec{\Gamma\Lambda}_1|$ given by the momentum of those excitons that condense in valleys 1 and 4, and similarly  $\Delta\varrho_{3,6}$ and $\Delta\varrho_{5,2}$ display an analogous modulation along directions
$\Vec{\Gamma\Lambda}_3$ and $\Vec{\Gamma\Lambda}_5$
with phase shifts $\varphi_3$ and $\varphi_5$,
respectively.
Here $\psi_{\Gamma}$
and $\psi_{\Lambda_1}$ are the periodic envelopes of Bloch states respectively at $\Gamma$ and $\Lambda_1$,   $\psi_{\Lambda_4}=\psi^*_{\Lambda_1}$, and the spin has been factored out, since the lattice space group contains a center of inversion and a unique $z$ axis   \cite{Mattheiss1973}.
It is clear that the total amount of charge displaced from the pristine background, as well as the amplitude of the charge modulation, are both driven by the condensate through $\sum_{\text{\bf k}}\,u_{\text{\bf k}}^0v_{\text{\bf k}}^0$.

Importantly, 
the arbitrariness of the phases
$\varphi_1$, $\varphi_3$, and $\varphi_5$ points to
a huge, continuous degeneracy of the ground state. Since the effect of any given two arbitrary phases is merely to rigidly shift the charge pattern $\Delta\varrho$ with respect to the frame origin 
(Methods), in the following we
take $\varphi_1=\varphi_3=\varphi_5=0$. The resulting density wave is slightly distorted in the generic case, in which all three phases take arbitrary values (see discussion below).

Figure 4C shows the overlap charge density of the envelopes obtained from first principles, $\sum_{\sigma}\psi^*_{\Gamma\sigma}(\text{\bf r})\,\psi_{\Lambda_1\sigma}(\text{\bf r})\;+\;\text{c.c.}$, which is proportional to $\Delta\varrho_{1,4}$
in the unit cell at the origin (we have added the subscript $\sigma$ to $\psi$ since the numerical envelopes are generically spinors in the presence of spin-orbit coupling). The density wave shows an asymmetric pattern---transferring charge mainly between the two Mo atoms, which breaks the inversion symmetry with respect to the origin of the cell [the red (blue) contour map points to a surplus (deficit) of charge]. 
This charge tranfer sets a local electric dipole with an in-plane texture, ${\bf P}_{1,4}(x,y)$, as $\Delta\varrho_{1,4}$ is modulated by 
$\exp{\![i(\Vec{\Gamma\Lambda}_1 \!\cdot \!\text{\bf r})]}$. 
This dipole may be regarded as the polarization of the excitons coherently built in the condensate \cite{Portengen1996b}.
Since the contributions to the  
dipole due to the remaining valleys, 
${\bf P}_{3,6}$ and ${\bf P}_{5,2}$, 
are obtained
by rotating ${\bf P}_{1,4}$  by respectively $2\pi/3$ and $-2\pi/3$ along the $z$ axis, the total
dipole ${\bf P}={\bf P}_{1,4}+{\bf P}_{3,6}+{\bf P}_{5,2}$ is parallel to the $z$ axis. We evaluate
this parallel component, $P_z$, through direct integration over the unit cell (Fig.~4D and Methods).

The overall charge pattern, $P_z(x,y)$, exhibits an anti-ferroelectric texture that breaks inversion symmetry.
This is shown in Fig.~4B, where local 
dipoles, which point out of the plane, are depicted as red arrows having length proportional to $|P_z|$. 
The electric dipole averages to zero over the unitary cell of the superstructure (solid frame),
which contains 72 atoms [with $\Gamma\Lambda\approx 2\pi/(3a)$] against 6 of the original cell (dashed frame).
The reconstructed Brillouin zone, which is again hexagonal in the plane but rotated by $\pi/6$ (SI Appendix, Fig.~S6D), is spanned by any two independent vectors chosen among the
$\Vec{\Gamma\Lambda}_i$'s.
In the generic, degenerate case that $\varphi_1$, $\varphi_3$, and $\varphi_5$ take arbitrary values, we expect a reduction of the maximum local value
of $|P_z|$ up to $2/3$, together with a variable tilt of the dipole in the plane.    

\begin{figure*}[htbp]
\centering
%\setlength{\unitlength}{1 cm}
%\begin{picture}(16,8)
%\put(0.2,-0.5){\includegraphics[trim=2.5cm 0 2cm 0cm,clip=true,width=7.5cm]{./N_01_HD.jpg}}
%\put(9.1,0.2){\includegraphics[trim=5cm 1.5cm 0 0,clip=true,width=7.5cm]{./Fig1b_new.pdf}}
%\includegraphics[trim=2.5cm 0 2cm 0cm,clip=true,width=5.5cm]{./N_01_HD.jpg}
%\qquad\qquad
%\put(0.0,-1.2){\includegraphics[trim=0cm 0cm 0 0,clip=true,width=16.0cm]{./Fig5.pdf}}
\includegraphics[width=11.4cm]{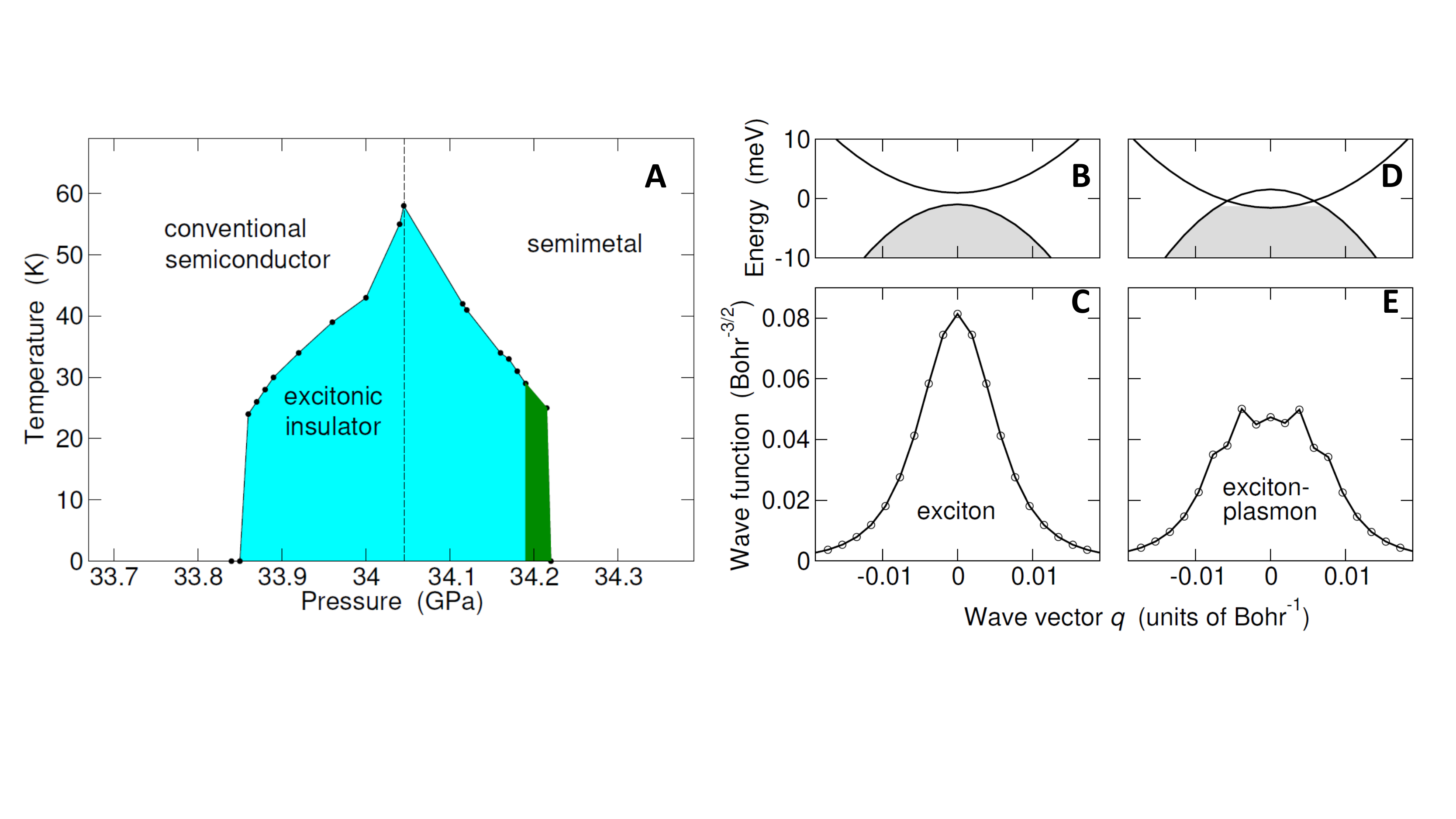}
%\put(0.5,7.5){\bf{a}}
%\put(0.5,3.0){\bf{b}}
%\put(4.2,7.5){\bf{c}}
%\put(4.2,3.0){\bf{d}}
%\put(10.2,5.9){\bf{e}}
%\end{picture}
\caption{
{\bf Excitonic insulator phase diagram. }
({\bf A})  Phase diagram in the $P$--$T$ space. Lines are guides to the eye. The shadowed area highlighted in cyan (green) colour is the excitonic gapped (gapless) phase. The vertical dashed line points to the semiconductor-semimetal boundary in the absence of excitonic effects.
({\bf B} to {\bf E}) Bare energy bands and wave function of the exciton driving the instability in the $e$-$h$ center-of-mass frame, evaluated in reciprocal space along the $\Gamma\Lambda$ direction. The $e$-$h$ pair of wave vector $q$ is made of a hole with momentum $-q$ and an electron with momentum $q + \Gamma\Lambda$ (in panels B and D the bare conduction band has been displaced by the vector $-\Vec{\Gamma\Lambda}$ and the shadowed region highlights occupied states). Going from $P=34$ GPa (panels B and C) to $P=34.12$ GPa (panels D and E), a Fermi surface forms as conduction and valence band overlap in energy. Consequently, plasmonic features appear in the exciton
wave function, the spectral weight accumulating close to the Fermi surface (panel E). In panel d the Fermi energy is negative as a consequence of the six-fold valley degeneracy at $\Lambda$.}
\end{figure*}

\section*{Semiconductor-semimetal crossover} 

The formation of a Fermi surface, made of six $e$ pockets in the $\Lambda$ valleys and one $h$ pocket at $\Gamma$, signals the transition from the semiconductor (Fig.~5B) to the semimetal (Fig.~5D) occurring in the absence of excitonic effects. Figures 5B and 5D show one of the conduction valleys, displaced by $-\Vec{\Gamma\Lambda}$ in {\bf k} space, and the valence band, the filled states being shadowed by gray colour. As the free carriers populating the Fermi pockets effectively screen the $e$-$h$ attraction, we replace the long-range Coulomb force $W$ in Eq.~3 with the vertex interaction proposed by Kozlov and Maksimov \cite{Kozlov1965} to establish self-consistently the range of the force; besides, we
extrapolate $P$-dependent masses from first principles (Methods). 

The resulting EI phase extends over a narrow interval of $\approx 0.35$ GPa, reaching a maximum critical temperature of $T \approx 60$ K at $P\approx 34.05$ GPa, which is the semiconductor-semimetal boundary in the normal state (vertical dashed line in Fig.~5A).  
Importantly, the downward shift of the valence band shown in Fig.~4A
opens / widens the gap over a pressure range that extends to values that would lead to a semimetal for $\left|\zeta_{i\text{\bf k}}\right| = 0$. 
In the $P-T$ diagram of Fig.~5A, the gapped excitonic phase, highlighted as a shadowed cyan area, is the overwhelming part of the larger region that sustains a finite condensate of excitons, $\left|\zeta_{i\text{\bf k}}\right|>0$. The remaining excitonic region---the thin green slice located between $P\sim$ 34.19 and 43.22 GPa---is gapless (SI Appendix, Fig.~S7) and ends on the semimetal frontier where $\left|\zeta_{i\text{\bf k}}\right|=0$. 
Here the critical pressure is equivalent to an amount of free carriers (the density per species is $1.1\cdot 10^{-7}$ Bohr$^{-3}$) comparable to the maximum number of excitons in the condensate ($2.2\cdot 10^{-7}$ Bohr$^{-3}$). 
This overall behaviour is in stark contrast with that of the EI candidate TiSe$_2$, which has a multivalley structure similar to that of MoS$_2$ but remains a semimetal due to the unintentional doping of Ti atoms \cite{Monney2009}.

The exciton responsible for the instability of the conventional semiconductor exhibits a mixed transverse--longitudinal polarization \cite{Knox1963}, due to the small $C_2$ symmetry of the $\Gamma\Lambda$ line (this is also the case of the displacement vectors of the vibrational mode of Fig.~6C). As one moves from the semiconductor to the semimetal, the exciton smoothly turns into a plasmon \cite{Kohn1967b},
as illustrated by the wave function in the $e$-$h$ center-of-mass frame (Methods). Whereas in the semiconductor (Fig.~5C) the amplitude is Lorentzian-like in {\bf k} space, similar to that of a familiar Wannier exciton in the bulk, in the semimetal it acquires plasmonic features, as the wave function accumulates close to the Fermi surface (Fig.~5E). Outside the EI phase, this exciton-plasmon dissolves into the continuum of $e$-$h$ excitations. Note that there may be other long-lived interband plasmons, since small gaps open in the $e$-$h$ energy continuum due to the degeneracy of $\Lambda$ valleys. Were there only one valley, then the Fermi energy would be at the crossing of $a$ and $b$ bands (ignoring the mass anisotropy, cf.~Fig.~5D) and the $e$-$h$ excitation spectrum would be gapless.
 
\begin{figure*}[htbp]
\centering
%\setlength{\unitlength}{1 cm}
%\begin{picture}(16,8.5)
%\put(0.2,-0.5){\includegraphics[trim=2.5cm 0 2cm 0cm,clip=true,width=7.5cm]{./N_01_HD.jpg}}
%\put(9.1,0.2){\includegraphics[trim=5cm 1.5cm 0 0,clip=true,width=7.5cm]{./Fig1b_new.pdf}}
%\includegraphics[trim=2.5cm 0 2cm 0cm,clip=true,width=5.5cm]{./N_01_HD.jpg}
%\qquad\qquad
%\put(0.0,-1.0){\includegraphics[trim=0cm 0cm 0 0,clip=true,width=16.0cm]{./Fig6.pdf}}
\includegraphics[trim=0cm 4cm 0 8 cm,clip=true,width=14.8cm]{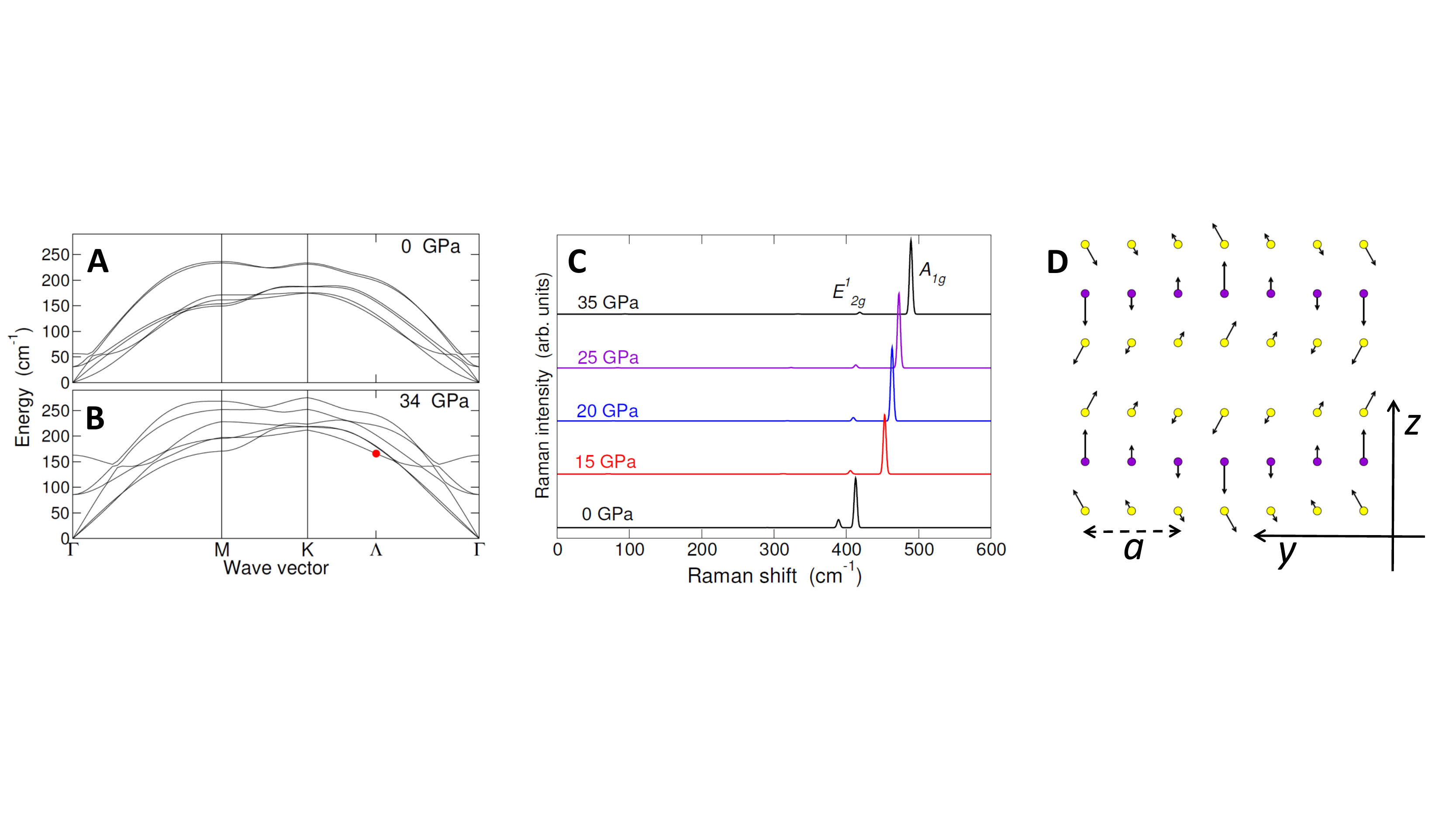}
%\put(0.5,7.5){\bf{a}}
%\put(0.5,3.0){\bf{b}}
%\put(4.2,7.5){\bf{c}}
%\put(4.2,3.0){\bf{d}}
%\put(10.2,5.9){\bf{e}}
%\end{picture}
\caption{
{\bf Phonon dispersion and Raman fingerprint. }
({\bf A} and {\bf B})  Dispersion of the lowest-energy phonon modes for $P=0$ (panel A) and 34 GPa (panel B), respectively, computed from first principles.
All modes harden with $P$. The red dot points to the lowest optical mode that is folded from $\Lambda$ into $\Gamma$ through the excitonic insulator phase transition.
({\bf C}) Raman spectrum of the normal phase from first principles, for pressures $P=$ 0, 15, 20, 25, 35 GPa, respectively from bottom to top. The peaks are broadened using Gaussians with a standard deviation of 2 cm$^{-1}$. The bright peak at lower (higher) frequency has $E_{2g}^1$ ($A_{1g}$) symmetry. The plot compares with Fig.~4(b) of Ref.~\cite{Cao2018}.   
({\bf D}) Displacement vectors for the mode labeled as a red dot in panel B, as viewed in the excitonic insulator reconstructed cell along the $\Gamma\Lambda$ direction [parallel to the $y$ axis in the adopted frame \cite{Mattheiss1973}]. The superlattice constant is $3a$. The violet (yellow) colour labels Mo (S) atoms. This mode is Raman-active and degenerate with the one folded from $\Lambda'$ to $\Gamma$.  
}
\end{figure*}

\section*{Raman fingerprint} Were ion displacements responsible for the building of electric dipoles in place of excitons, the frequency of the phonon of momentum {\bf q} $=\Vec{\Gamma\Lambda}$ and consistent symmetry would soften (or at list exhibit a dip) at the onset of the new phase \cite{Gruner2018}. The phonon dispersion obtained from first principles, respectively at $P=0$ (Fig.~6A) and 34 GPa (Fig.~6B), shows the opposite behaviour, with all low-energy modes hardening with $P$ (Methods and SI Appendix, Fig.~S8 for the 2$H_a$ phase). Therefore, the anti-ferroelectricity has a purely electronic origin.
This prediction is consistent with recent diffraction measurements, which ruled out any periodic lattice distortion above 40 Kelvin \cite{Goncharov2020}.

The evolution of Raman spectrum with pressure, as obtained from first principles in Fig.~6C (structure 2$H_c$) and SI Appendix, Fig.~S9 (2$H_a$), compares with observed data with the exception of the $E'$ peak at 174 cm$^{-1}$
[Fig.~4(b) of Ref.~\cite{Cao2018}], which appears below 150 K and above 30 GPa but is missed by the theory for the normal phase. Cao and coworkers proposed \cite{Cao2018} this mode is a transverse acoustic phonon of finite momentum, which becomes bright at the onset of a charge density wave, due to the reconstruction of the Brillouin zone.   
Whereas the first-principles spectrum for the excitonic phase is presently out of reach, below we confirm the essence of Cao's explanation by identifying $E'$ as the lowest optical phonon at $\Lambda$. This is the fingerprint of the anti-ferroelectric charge density wave associated with exciton condensation.

The symmetry group of the anti-ferroelectric ground state depicted in Fig.~4B only includes the identity operation. Therefore, all 216 vibrational modes are in principle infrared and/or Raman active. However, since the EI critical temperature is relatively low and the $E'$ peak is extremely bright, we expect that the new mode is an optical phonon of momentum $\Lambda$, which is Raman active through the folding into the zone center and
strongly couples with $\text{\bf P}$. 
Since $\text{\bf P}(x,y)$ originates everywhere in the cell from the inter-layer
vertical displacement of the charge between two neighbour Mo atoms, it will mainly couple with those optical oscillations of Mo atoms that occur along the $z$ axis. In fact, these vibrations linearly change the Mo-Mo distance and hence the local dipole strength, whereas the amount of displaced charge, which is ruled by the long-range part of Coulomb interaction, changes weakly with the oscillation. From direct inspection of phonon eigenvectors,
there is one candidate only below 400 cm$^{-1}$, i.e., the lowest optical mode of frequency 164 cm$^{-1}$ located at $\Lambda$, which is highlighted by a red dot in Fig.~6B. As shown by the displacement vectors in the EI reconstructed cell displayed in Fig.~6D, the Mo atoms oscillate out of phase along the $z$ direction with an in-plane modulation of period $3a$ along the $\Vec{\Gamma\Lambda}$ direction (parallel to the vertical axis of Fig.~4B), hence matching the periodicity of $P_z(x,y)$ in the plane.
This superlattice vibration is twice degenerate, due to the additional folding of the phonon with independent 
wave vector $\Vec{\Gamma\Lambda}'$.
Note that the observed intensity of the $E'$ mode is constant up to 60 K, which compares with the EI critical temperature. 
In summary, the $E'$ mode points to the excitonic insulator in the $P-T$ space.

\section*{Discussion} 

Both $2H_c$ and $2H_a$ phases coexist \cite{Nayak2014,Chi2014,Bandaru2014,Chi2018} in the region of visibility of the $E'$ mode, which extends between 30 and 50 GPa at 5 Kelvin \cite{Cao2018}. The lower bound agrees with our prediction, since in the $2H_a$ structure the EI sets in at $P\sim 28$ GPa (SI Appendix, Fig.~S4) with a mode frequency of 166 cm$^{-1}$ (SI Appendix, Fig.~S8). The upper bound of 50 GPa is larger than our expectation of $\sim 34$ GPa for the $2H_c$ phase. However, recent diffraction measurements on single crystals \cite{Goncharov2020}, though only available at temperatures higher than 40 Kelvin, suggest that the critical upper pressure could be actually much lower, being artificially enhanced in powders due to the deviatoric stress field applied to randomly oriented crystallites.
 
In addition, other Raman features unexplained so far \cite{Cao2018} point to the EI scenario: (i) the observation of modes supposedly forbidden or silent (ii) the anomalous frequency variation of the out-of-plane $A_{1g}$ mode accompanying the onset of the $E'$ mode. Since the understanding of the available electrical transport measurements \cite{Chi2014,Cao2018} is complicated by the mixture of phases in the high-pressure cell, we do not speculate on the origin of the resistivity peak that was tentatively assigned \cite{Chi2014} to the EI.

The huge degeneracy of the EI ground state, associated with condensate phases $\varphi_1$, $\varphi_3$,
and $\varphi_5$, points to the emergence of acoustic-like electronic excitations---collective phase modes that, if gapless, would manifest exciton superfluidity \cite{Kohn1967b}. Within the two-band model of an isotropic semimetal, Kozlov and
Maksimov \cite{Kozlov1965b} predicted that the ``excitonic sound'' velocity, $c_{\text{exciton}} =\hbar k_F (3 m_a m_b)^{-1/2} $, is proportional to Fermi wave vector in the normal phase, $k_F$ ($m_a$ and $m_b$ are valence and conduction band masses). By taking average values at the
EI / semimetal boundary, we estimate
$c_{\text{exciton}} \sim 2\cdot 10^4$ m/s, which is much higher than the sound velocity of the stiffest acoustic phonon branch (Fig.~6B), $c_{\text{phonon}} \sim 8\cdot 10^3$ m/s. Therefore, the phase mode of the exciton condensate should be experimentally accessible.

\section*{Conclusion}

In summary, we have demonstrated that a real excitonic insulator phase sets in between the semiconducting and semimetallic phases of MoS$_2$, building on calculations from first principles and available spectroscopic data.
These findings call for further investigation of some fascinating possibilities. A first question is the manifestation of the macroscopic quantum coherence of the exciton condensate, which might occur through the observation of low-lying collective modes associated with the oscillation of the condensate phase $\varphi(${\bf r}$,t)$. Another issue is whether the superconductivity observed above 90 GPa is related to the excitonic phase, as the overscreening action of surviving exciton-plasmons might act as unconventional glue for Cooper pairs. We hope our study may stimulate further work along these paths.

\begin{methods}

\subsection{Computational details of ground-state calculation from first principles}

The lattice parameters and the ground-state electronic structure for the three values of  pressure were obtained within density functional theory (DFT), with a plane wave basis set as implemented in the Quantum ESPRESSO package \cite{Giannozzi2009,Giannozzi2017}, using the generalized gradient approximation Perdew-Burke-Ernzerhof (PBE) parametrization \cite{pbe96}. A kinetic energy cutoff of 100 Ry was adopted for the wave functions, and fully relativistic norm-conserving pseudopotentials \cite{hamann2013optimized} were used to take into account spin-orbit interaction. Van der Waals interactions, included by using the Grimme approximation method, were found to be relevant only at zero pressure, as already shown in Ref.~\cite{tosatti}. 

\subsection{Phonons} 

Phonon dispersions were calculated by using a Density Functional Perturbation Theory approach \cite{Baroni2001}. We used a 10 $\times$ 10 $\times$ 3 Monkhorst-Pack grid for the integration in the Brillouin zone; the dynamical matrix at a given point of the Brillouin zone was obtained from a Fourier interpolation of the dynamical matrices computed on a 5 $\times$ 5 $\times$ 1 {\bf q}-point mesh.

\subsection{Quasiparticles and excitons} 

Many-body calculations \cite{Onida-Reining-Rubio_2002,hybertsen1986,strinati1988g}  were performed by using the Yambo code \cite{Marini2009,yambo}. Quasiparticle corrections to the Kohn-Sham energies were evaluated using the $G^0W^0$ approximation for the self-energy, the dynamical dielectric screening been accounted for within the plasmon-pole approximation \cite{godby}.  To speed-up the convergence of quasiparticle energies with respect to the number of empty bands in the sum over states occurring 
in the calculation of the polarizability and self energy, we have adopted the scheme proposed in Ref.~\cite{bruneval2008accurate}.
Fifty empty bands were used to build the polarizability and to integrate the self-energy (SI Appendix, Fig.~S10); the Brillouin zone was sampled by using a 27 $\times$ 27 $\times$ 3 {\bf k}-point grid. Quasiparticle energies were converged by using 68 Ry and 15 Ry kinetic energy cutoffs for the exchange and correlation parts of the self-energy (SI Appendix, Fig.~S11), respectively.
Excitation energies and dispersion of the lowest exciton with finite wavevector {\bf q} were calculated by solving the Bethe-Salpeter equation (BSE) using a developer's version of the Yambo code where the finite-{\bf q} BSE was implemented as described in Refs.~\cite{gatti2013,Shirley2000}.
The static screening in the direct term was calculated within the random phase approximation with inclusion of local field effects; the Tamm-Dancoff approximation for the Bethe-Salpeter Hamiltonian was employed, after having verified that the correction introduced by coupling the resonant and antiresonant part was negligible for {\bf q} = 0. Converged excitation energies were obtained considering respectively 3 valence and 5 conduction bands in the Bethe-Salpeter matrix, the irreducible Brillouin zone being sampled with a 27 $\times$ 27 $\times$ 3 {\bf k}-point grid (SI Appendix, Fig.~S12). 

\subsection{Computational details of the two-band model} 

The effective-mass framework builds on the knowledge of conduction  
\begin{equation}
\varepsilon_b(\text{\bf k})= \frac{G}{2} + 
\frac{\hbar^2}{2}\left[\frac{(k_{\parallel}+\Gamma\Lambda)^2}{m_{b\parallel}} +
\frac{k_{\perp}^2}{m_{b\perp}}+
\frac{k_{z}^2}{m_{bz}}\right]
\end{equation}
and valence 
\begin{equation}
\varepsilon_a(\text{\bf k})= -\frac{G}{2} - 
\frac{\hbar^2}{2}\left[\frac{k_{\parallel}^2}{m_{a\parallel}} +
\frac{k_{\perp}^2}{m_{a\perp}}+
\frac{k_{z}^2}{m_{az}}\right]
\end{equation}
energy bands. Here $G>0$ ($G<0$) is the indirect bandgap (band overlap) for pressures below (above)
the semiconductor-semimetal threshold---in the absence of excitonic effects---and the momentum components, $k_{\parallel}$, $k_{\perp}$, $k_z$, are projected along the principal axes of the effective mass tensor \cite{Pikus1974}, the corresponding masses being $m_{i\parallel}$, $m_{i\perp}$,
$m_{iz}$, with $i=a,b$. These axes are respectively parallel ($k_{\parallel}$) and perpendicular [in- ($k_{\perp}$) and out-of-plane ($k_z$)] to the $\Vec{\Gamma\Lambda}$ direction, the axis origin being placed at the band edge. We emphasize that all parameters of the two-band model, for a given pressure, are fixed and obtained from first principles. In particular, the bandgap and the effective masses are extracted from GW bands, 
as illustrated in Figs.~2D to F, and hence include the mean-field renormalization due to $e$-$e$ interactions. The (modulus) of the screened $e$-$h$ Coulomb attraction in momentum space,  
\begin{equation}
    W(\text{\bf q})=\frac{1}{\kappa_r}\frac{4 \pi e^2}{\Omega}\frac{1}{q^2},
\end{equation}
depends on the static dielectric constant, $\kappa_r$, which is obtained as the inverse of the first-principles dielectric tensor, $1/[\epsilon^{-1}(\mathbf{q}=0)]_{\mathbf{G}=\mathbf{G'}=0}$, in the long-wavelength, macroscopic limit, as illustrated in Fig.~3B (here $\Omega$ is the crystal volume and {\bf G} the reciprocal lattice vector). 

In the semimetal, the $P$-dependent values of $G$,
$m_{i\parallel}$, $m_{i\perp}$,
$m_{iz}$, and $\kappa_r$ are derived as linear extrapolations of first-principles data at $P=$ 25 and 34 GPa, respectively. Since free $e$ and $h$ carriers effectively screen the interaction by
adding a metal-like, intraband contribution to the polarizability, we modify the dressed Coulomb potential  as 
\begin{equation}
    W(\text{\bf q})=\frac{1}{\left[\kappa_r + 4\pi e^2 {\cal{D}}(\varepsilon_{\text{F}}) / q^2\right]}\frac{4 \pi e^2}{\Omega}\frac{1}{q^2}.
\end{equation}
Here the Thomas-Fermi term, proportional to the density of states, ${\cal{D}}(\varepsilon)$, evaluated at the Fermi energy, $\varepsilon_{\text{F}}$,  
removes the long-wavelength divergence of $W$. We obtain numerically ${\cal{D}}$ through the summation of localized Gaussian functions over a fine grid in {\bf k} space, as well as
$\varepsilon_{\text{F}}$ by imposing overall charge neutrality  
(we take into account the six-fold degeneracy of conduction band).

\subsection{Two-band Bethe-Salpeter equation} 

In the semiconductor, the exciton wave function is
\begin{equation}
\left|\text{exciton}\right> = \sum_{\text{\bf k}} \phi_{\text{\bf k}}  \,\hat{b}^+_{\text{\bf k}}\hat{a}_{\text{\bf k}} \left|\Phi_0\right>,   
\end{equation}
where $\phi_{\text{\bf k}}$ is the probability amplitude of a bound $e$-$h$ pair in momentum space.
The Bethe-Salpeter equation of motion for $\phi_{\text{\bf k}}$ is
\begin{equation}
\left[\varepsilon_b({\text{\bf k}})-\varepsilon_a({\text{\bf k}})\right]\phi_{\text{\bf k}}-\sum_{\text{\bf k}'}
W(\text{\bf k}-\text{\bf k}')   \phi_{\text{\bf k}'}\;=\varepsilon_{\text{exc}}\,\phi_{\text{\bf k}},
\end{equation}
where $\varepsilon_{\text{exc}}$ is the excitation energy of the exciton, whose negative value signals the instability.
We solve this equation by numerical discretization in {\bf k} space and assess convergence by refining the mesh as well as varying the momentum cutoff. Note that the singularity of Coulomb potential for $\left|\mathbf{q}\right| \rightarrow 0$ is harmless, as we integrate $W$ over a small parallelepiped, in a semi-analytical, accurate manner. We have benchmarked the convergence of our calculations against known analytical or high-precision results, as shown for bulk Wannier excitons in SI Appendix, Fig.~S13 and for anisotropic excitons with a well-defined azimuthal quantum number \cite{Pedersen2016} in SI Appendix, Fig.~S14. 

In the semimetal ground state, a small area of {\bf k} space around the origin is populated by electrons in band $b$ and holes in band $a$. In addition, due to band anisotropy \cite{Zittartz1967}, in narrow regions nearby there are either electrons or holes only, which prevents from exciting $e$-$h$ pairs due to Pauli exclusion principle. Therefore, the Bethe-Salpeter equation of motion must be modified as \cite{Kohn1967b} 
\begin{equation}
\left[\varepsilon_b({\text{\bf k}})-\varepsilon_a({\text{\bf k}})\right]\phi_{\text{\bf k}}-\sum_{\text{\bf k}'}
W(\text{\bf k}-\text{\bf k}')\left[n_a(\text{\bf k}')-n_b(\text{\bf k}')\right]  \phi_{\text{\bf k}'}\;=\varepsilon_{\text{exc}}\,\phi_{\text{\bf k}},
\end{equation}
where $n_i(\text{\bf k})$ is the occupancy factor of the $i$th band in the normal ground state, which takes value either 0 or 1. The `counting' prefactor of $W$, $\left[n_a-n_b\right]$, removes scattering channels forbidden by Pauli blocking and is responsible of the plasmon-like features shown Fig.~5E. Note that in the semiconductor, $n_a(\text{\bf k})=1$
and $n_b(\text{\bf k})=0$, hence one regains the standard form of equation 11.

\subsection{Self-consistent theory of the excitonic insulator within the two-band model}

The EI bands (circles in Fig.~4A) are 
$E_{b\text{\bf k}}= \left[\varepsilon_b({\text{\bf k}})+ \varepsilon_a({\text{\bf k}})\right]/2 + E_{\text{\bf k}}$ and $E_{a\text{\bf k}}= \left[\varepsilon_b({\text{\bf k}})+\varepsilon_a({\text{\bf k}})\right]/2 - E_{\text{\bf k}}$, with $E_{\text{\bf k}}$ being fixed by the solution of the gap equation 3 of main text for $\Delta^0_{\text{{\bf k}}}$ (through $\zeta^0_{\text{\bf k}}$). Equation 3 of main text is solved self-consistently by means of numerical recursion, exploiting the exciton wave function $\phi_{\text{\bf k}}$ as a seed \cite{Varsano-Rontani2017}. If the semimetal is the normal ground state, the gap equation maintains the form 3 of main text, provided that: (i) The summation over {\bf k}$'$ is limited to those points whose occupancies are such that $\left[n_a(\text{\bf k}')-n_b(\text{\bf k}')\right]\neq 0$ to comply with Fermi statistics \cite{Zittartz1967}. (ii) The dressed Coulomb interaction $W$ is renormalized by a vertex correction associated with the EI ground state \cite{Kozlov1965}, as the opening of the many-body gap significantly enhances the $e$-$h$ attraction---by suppressing screening---with respect to the gapless normal phase. Therefore, following Kozlov and Maksimov \cite{Kozlov1965}, for small momentum transfer $q$ the dressed interaction $W$ appearing in equation 3 of main text takes the self-consistent form
\begin{equation}
  W(\text{\bf q})=\frac{1}{\left[1 + \alpha /(\Delta^0_{\text{{\bf k}}_{\text{F}}})^2\right]}\frac{1}{\kappa_r}\frac{4 \pi e^2}{\Omega}\frac{1}{q^2},  
\end{equation}
where the gap function at the Fermi surface, $\Delta^0_{\text{{\bf k}}_{\text{F}}}$, which is determined recursively, removes the long-wavelength divergence as one approaches the EI-semimetal boundary. Here $\Delta^0_{\text{{\bf k}}_{\text{F}}}$ is an average value defined as $\Delta^0_{\text{{\bf k}}_{\text{F}}}=\left[\Delta^0_{k_{x\text{F}},0,0}\Delta^0_{0,k_{y\text{F}},0}\Delta^0_{0,0,k_{z\text{F}}}\right]^{1/3} $, with 
$k_{x\text{F}}$ given implicitly by $\varepsilon_{\text{F}}= \varepsilon_b(k_{x\text{F}},0,0)$, and similarly for
$k_{y\text{F}}$ and $k_{z\text{F}}$. The constant $\alpha$, for given band overlap $G<0$, is $\alpha=[\left|G_0\right|
(\varepsilon_{\text{F}}-G/2)^{3/2}]^{1/2}$, where $\left|G_0\right|=9.38$ meV is the maximum magnitude of the band overlap at which $e$-$h$ pairing takes place. We neglect the modification of Eq. 13 for large momentum transfer, as it turns out to be irrelevant numerically. Whereas the vertex form 13 was originally proposed \cite{Kozlov1965} for the case of spherically symmetric $e$ and $h$ pockets, we notice that, at the semiconductor-semimetal threshold, the exciton responsible for the instability is essentially isotropic (SI Appendix, Fig.~S5). At finite temperature, $T$, the gap equation takes the form
\begin{equation}
2E_{\text{\bf k}}\,\zeta^0_{\text{\bf k}}-\sum_{\text{\bf k}'}
W(\text{\bf k}-\text{\bf k}')\,\zeta^0_{\text{\bf k}'}\left[f_{\text{F}}(E_{a\text{\bf k}'}-\varepsilon_{\text{F}})
-f_{\text{F}}(E_{b\text{\bf k}'}-\varepsilon_{\text{F}})
\right] 
\;=\;0,
\end{equation}
where $f_{\text{F}}(x)=1/[1 + \exp{(\beta x})]$ is Fermi distribution function, with $\beta = 1/k_{\text B}T$ and
$k_{\text B}$ being Boltzmann constant, and we neglect the small renormalization of the chemical potential due to the presence of the exciton condensate.

\subsection{Multivalley band structure}

The calculation of the EI band structure relies on the theory by Monney and coworkers \cite{Monney2009} to include valley degeneracy. This approach, based on Green functions, generalizes to multiple bands the original theory by J{\'e}rome and coworkers \cite{Kohn1967}. For every
{\bf k} point, the EI band energies (solid lines in Fig.~4A and SI Appendix, S7) are found as the seven roots of the equation  
\begin{equation}
z - \varepsilon_{a}(\text{\bf k}) - \sum_{i=1}^6 \frac{\left| \Delta_i(\text{\bf k})\right|^2}{ z - \varepsilon_{ib}(\text{\bf k}) } = 0
\end{equation}
[cf.~Eq.~(8) of Ref.~\cite{Monney2009}],
after the magnitudes of the  excitonic gap components, $\Delta_i(\text{\bf k})$, are obtained as follows. The gap function is defined as
\begin{equation}
\Delta_i(\text{\bf p})=\sum_{\text{\bf k}}W\!(\text{\bf k})\,\zeta_{i\text{\bf k + p}},
\end{equation}
with $\zeta_{i\text{\bf k}}$, apart from a phase factor, being the equal-time interband excitonic coherence 
$F^{\dagger}_i\!(\text{\bf k},t,t)$ defined in Eq.~(4)
of Ref.~\cite{Monney2009},
\begin{equation}
\zeta_{i\text{\bf k}} = -iF^{\dagger}_i\!(\text{\bf k},t+\delta,t)=  \frac{1}{2\pi i}\int_{-\infty}^{\infty}
\!\!\!\text{d}\omega\, F^{\dagger}_i\!(\text{\bf k},\omega)
\,\text{e}^{-i\omega\delta},
\end{equation}
and $\delta\rightarrow 0^+$ being a positive infinitesimal quantity. The integral 17 is evaluated through contour integration, the Fourier transform $F^{\dagger}_i\!(\text{\bf k},\omega)$ being derived from the equations of motion of Green functions \cite{Monney2009} as
\begin{eqnarray}
F^{\dagger}_i\!(\text{\bf k},\omega)  = 
- \Delta_i(\text{\bf k})
\left[   
\omega - \varepsilon_{a}(\text{\bf k}) - \sum_{j\neq i} \frac{\left| \Delta_j(\text{\bf k})\right|^2}{ \omega - \varepsilon_{jb}(\text{\bf k}) }
\right]^{-1} \times \nonumber \\
\left[   
\omega - \varepsilon_{ib}(\text{\bf k}) - 
\left| \Delta_i(\text{\bf k})\right|^2 
\Big(
\omega - \varepsilon_{a}(\text{\bf k}) - \sum_{j\neq i}
\frac{\left| \Delta_j(\text{\bf k})\right|^2}
{\omega - \varepsilon_{jb}(\text{\bf k})}
\Big)^{-1}
\right]^{-1}.
\end{eqnarray}
Whereas this expression would generically lead to an intractable system of six coupled equations for the $\Delta_i$'s, we exploit the high symmetry of the problem  
to simplify the form of $F^{\dagger}_i\!(\text{\bf k},\omega)$ and recover a single gap equation. As discussed in the main text and SI Appendix, Fig.~S6, the symmetrizing effect of $e-h$ attraction
makes $\Delta^0_{\text{\bf k}}$ almost independent from the azimuthal angle
$\varphi_{\text{\bf k}}$, with {\bf k} $\equiv (k,\varphi_{\text{\bf k}},k_z)$ being expressed in cylindrical coordinates ($k$ is the in-plane radial distance and $k_z$ the component along the $z$ axis).
Therefore, it is natural to assume that $\Delta_i$ has cylindrical symmetry, 
$\Delta_i(\text{\bf k})=
\Delta(k,k_z)\text{e}^{i\varphi_i}$. Since we are mainly
interested in the region $\text{\bf k}\approx 0$, we also neglect 
the azimuthal dependence of
$\varepsilon_{ib}(\text{\bf k})$ in the denominator of $F^{\dagger}_i$, obtaining
\begin{equation}
F^{\dagger}_i\!(\omega)  = 
-\frac 
{\Delta_i}
{\left(   
\omega - \varepsilon_{a} - \frac{5\left| \Delta\right|^2}{ \omega - \varepsilon_{ib} }
 \right)
\left(   
\omega - \varepsilon_{ib} -
\frac{
\left| \Delta\right|^2} 
{
\omega - \varepsilon_{a} - 
\frac{5\left| \Delta\right|^2}
{\omega - \varepsilon_{ib}}
}
\right)},
\end{equation}
where we omitted the dependence of terms on {\bf k} in the notation.
Equation 19 is now easily integrated, giving a single
self-consistent gap equation.
This has the same form of the equation 3 of the two-band model, provided that 
$\Delta^0_{\text{\bf k}}$ is replaced
with $\sqrt{6}\,\Delta_i(\text{\bf k})$. 

\subsection{Ground state wave function}

The contour integration of equal-time Green functions provides us
with all interband coherences and band populations, i.e.,
$ \left<\Psi_{\text{EI}}\right| 
\hat{b}^+_{j\text{\bf k}} \hat{b}_{i\text{\bf k}} \left|\Psi_{\text{EI}}\right> =
\Delta_i^*\Delta_j/ 2E(E+\varepsilon_{b}/2-\varepsilon_{a}/2)$, $ \left<\Psi_{\text{EI}}\right|  
\hat{b}^+_{i\text{\bf k}}
\hat{b}_{i\text{\bf k}}
\left|\Psi_{\text{EI}}\right> = (v^0)^2/6$, $ \left<\Psi_{\text{EI}}\right|  
\hat{a}^+_{\text{\bf k}}
\hat{a}_{\text{\bf k}}
\left|\Psi_{\text{EI}}\right> = (u^0)^2$,
where we omitted the dependence of right-hand-side terms on {\bf k} to
ease the notation, neglected the in-plane anisotropy of valence band, $\varepsilon_{ib}=\varepsilon_{b}$, and put
$E =
\{[\varepsilon_b-\varepsilon_a]^2/4 + |\Delta^0|^2\}^{1/2}$. This allows us to write explicitly the ground state wave function,
\begin{equation}
\left|\Psi_{\text{EI}}\right> =
\prod_{\text{\bf k}}\hat{\gamma}^+_{\text{\bf k}}
\left|{\text{vacuum}}\right>,
\end{equation}
in terms of Bogoliubov-Valatin-like   creation operators, $\hat{\gamma}^+$, which are defined as
\begin{equation}
\hat{\gamma}^+_{\text{\bf k}}=u^0_{\text{\bf k}}\, \hat{a}^+_{\text{\bf k}}
+\frac{v^0_{\text{\bf k}}}{\sqrt{6}}\sum_{i=1}^6
\text{e}^{-i\varphi_i}\,\hat{b}^+_{i\text{\bf k}}.
\end{equation} 
As discussed in the main text, time reversal symmetry limits the number of independent condensate phases to three: $\varphi_1$, 
$\varphi_3$, and $\varphi_5$ (recall that $u^0_{\text{\bf k}}=u^0_{-\text{\bf k}}$ and $v^0_{\text{\bf k}}=v^0_{-\text{\bf k}}$
are real positive quantities; see SI Appendix, Fig.~S6C).

\subsection{Inversion symmetry breaking}

The ground state wave function allows us to understand the symmetry breaking associated with exciton condensation.  
The inversion operator, $\hat{\cal{I}}$, acts differently on 
$b_i$ and $a$ Bloch states, since the envelope function at
$\Gamma$ is odd:
$\hat{\cal{I}}\hat{a}^+_{\text{\bf k}}=-\hat{a}^+_{-\text{\bf k}}$, $\hat{\cal{I}}\hat{b}^+_{1\text{\bf k}}=\hat{b}^+_{4-\text{\bf k}}$, etc. Therefore, the inverted ground state, $\hat{\cal{I}}\left|\Psi_{\text{EI}}\right>$, is not proportional to the
original one:
\begin{eqnarray}
&&\left<\Psi_{\text{EI}}\right|
\hat{\cal{I}}
\left|\Psi_{\text{EI}}\right>  = \prod_{\text{\bf k}} 
{\Big\{}
-(u^0_{\text{\bf k}})^2
\nonumber\\
&+&\frac{(v^0_{\text{\bf k}})^2}{3}
\left[\cos{(2\varphi_1)}+\cos{(2\varphi_3)}+\cos{(2\varphi_5)}
\right]
\Big\}.
\end{eqnarray}
The magnitude of the expression enclosed in curly brackets is less than one (unless $\varphi_1=\varphi_3=\varphi_5=\pm \pi/2$, i.e., $\hat{\cal{I}}\left|\Psi_{\text{EI}}\right> = - \left|\Psi_{\text{EI}}\right>$), hence, in the thermodynamic limit, the
overlap between $\hat{\cal{I}}\left|\Psi_{\text{EI}}\right>$ and
$\left|\Psi_{\text{EI}}\right>$ tends to zero as the two states become
orthogonal. Since the ground state has a lower symmetry than the Hamiltonian, inversion symmetry is broken. 

\subsection{Charge density wave}

The form 5 of the purely electronic charge density wave, $\Delta\varrho=
\Delta\varrho_{1,4}+\Delta\varrho_{3,6}+\Delta\varrho_{5,2}$, 
is derived in a 
straightforward manner by averaging the density operator, $\hat{\varrho}(\text{\bf r})=2\,\hat{\psi}^{\dagger}\!(\text{\bf r})\,
\hat{\psi}(\text{\bf r})$, over $\left|\Psi_{\text{EI}}\right>$,
with the Fermi field operator, $\hat{\psi}(\text{\bf r})$, 
being defined as
\begin{equation}
\hat{\psi}(\text{\bf r})=\sum_{\text{\bf k}}{\text{e}}^{i\text{\bf k}
\cdot\text{\bf r}}
\left[\psi_{\Gamma}\!(\text{\bf r})\,
\hat{a}_{\text{\bf k}}
+ \sum_{i=1}^6 \psi_{\Lambda_i}\!(\text{\bf r})\,
\hat{b}_{i\text{\bf k}}
\right].
\end{equation}
Cross-terms proportional to
$\psi^*_{\Lambda_i}\psi_{\Lambda_j}$ average out to zero, once summed together,
as the various $\psi_{\Lambda_i}$'s are obtained one from the other by either rotation by $\pm 2\pi/3$ along the $z$ axis or complex conjugation.
Apart from the envelope functions, which have the lattice periodicity, $\Delta\varrho$ depends on
{\bf r} through a sum over three
exponentials, whose imaginary arguments 
are respectively (times the prefactor $i$)
$\Vec{\Gamma\Lambda}_1 \cdot \text{\bf r} - \varphi_1$, $\Vec{\Gamma\Lambda}_3 \!\cdot \!\text{\bf r} - \varphi_3$, and
$\Vec{\Gamma\Lambda}_5 \!\cdot \!\text{\bf r} - \varphi_5$, as illustrated in the main text.

We show below that, for any given two condensate phases, say $\varphi_1$ and $\varphi_3$, there exist a lattice vector
$\text{\bf R}_{\text{shift}}$ and a phase  
$\varphi_5 = -\varphi_1-\varphi_3 $ such that a rigid translation of the density wave by $\text{\bf R}_{\text{shift}}$ provides
the density wave corresponding to
$\varphi_1=\varphi_3=\varphi_5=0$, i.e.,
$[\Delta\varrho(\text{\bf r}-\text{\bf R}_{\text{shift}})]_{\varphi_1,\varphi_3,\varphi_5}=
[\Delta\varrho(\text{\bf r})]_{0,0,0}$.

Let us construct explicitly
$\text{\bf R}_{\text{shift}}$ as
$\text{\bf R}_{\text{shift}} = -  \text{\bf R}_{\parallel} - \text{\bf R}_{\perp}$, 
where $\text{\bf R}_{\parallel}= n_{\parallel} 
\text{\bf t}_2$ and $\text{\bf R}_{\perp}= n_{\perp} (2 \text{\bf t}_1 + \text{\bf t}_2)$ are respectively parallel and
perpendicular to $\Vec{\Gamma\Lambda}_1$ (SI Appendix, Fig.~S6C),
$n_{\parallel}$ and $n_{\perp}$ are integers
to be determined, and $\text{\bf t}_1$,
$\text{\bf t}_2$ are the primitive vectors that generate the hexagonal lattice
in Mattheiss' coordinate frame \cite{Mattheiss1973}. Since $\Vec{\Gamma\Lambda}_1$ is generically not
 commensurable with the reciprocal lattice vectors,  there exists an integer
 $n_{\parallel}$ 
 such that $\Vec{\Gamma\Lambda}_1 \cdot \text{\bf R}_{\parallel} = \varphi_1$ with arbitrary accuracy \cite{Kohn1967b}, modulus an integer multiple of $2\pi$. Similarly, we may fix
 $n_{\perp}$ such that
 $\Vec{\Gamma\Lambda}_3 \cdot \text{\bf R}_{\perp} = - \Vec{\Gamma\Lambda}_5 \cdot \text{\bf R}_{\perp}
 = \varphi_3 - \Vec{\Gamma\Lambda}_3 \cdot \text{\bf R}_{\parallel} 
 $. Finally, we take $\varphi_5 = - \varphi_3
 +2 \Vec{\Gamma\Lambda}_3 \cdot \text{\bf R}_{\parallel} =
 - \varphi_3 - \varphi_1 $. One may verify, by direct substitution into the expression
 $\Delta\varrho=
\Delta\varrho_{1,4}+\Delta\varrho_{3,6}+\Delta\varrho_{5,2}$, that $[\Delta\varrho(\text{\bf r}-\text{\bf R}_{\text{shift}})]_{\varphi_1,\varphi_3,-\varphi_1-\varphi_3}=
[\Delta\varrho(\text{\bf r})]_{0,0,0}$, qed.

This theorem implies that the
set of charge density waves $[\Delta\varrho(\text{\bf r})]_{0,0,\varphi_5}$ labeled by
the continuous parameter $\varphi_5$ spans all possible modulations of the electronic charge density of the EI, each realization having in turn a huge  
translational degeneracy, which is parametrized by the two continuous variables $\varphi_1$ and $\varphi_3$.

\subsection{Anti-ferroelectric order} 

The electronic charge density wave of the EI ground state (Eq.~5 of main text) induces an out-of-plane electric dipole, $P_z({\text{\bf R}}_i)$, in the $i$th cell of the pristine 2$H$ phase located at ${\text{\bf R}}_i$, with $i=1,\ldots,N$ ($N$ is the total number of cells). This is illustrated in Fig.~4B, where the dipoles $P_z({\text{\bf R}}_i)$ are depicted as red arrows. The local dipole $P_z({\text{\bf R}}_i)$ is given by the coherent superposition of
three density waves, whose characteristic wave vectors are $\text{\bf q}_i=\Vec{\Gamma\Lambda}_i$, with $i=1,3,5$,
\begin{equation}
P_z({\text{\bf R}}_i) = \frac{P_{z0}}{\Omega}\sum_{\text{\bf k}}\frac{4}{\sqrt{6}}
u_{\text{\bf k}}^0v_{\text{\bf k}}^0\!
\sum_{j=1,3,5}
\cos{(\text{\bf q}_j} \cdot {\text{\bf R}}_i).    
\end{equation}
The maximum value, $P_z(0)$, is shown in Fig.~4D.
Here $P_z({\text{\bf R}}_i)$ is evaluated within the envelope function approximation, the factor $P_{z0}$ being derived from first principles through 
the overlap charge density of the periodic part of conduction
and valence Bloch states at $\Gamma$ and $\Lambda$,
respectively, which is shown in Fig.~4C. The latter is numerically integrated over the pristine unit cell volume, $\Omega_{\text{cell}}$:
\begin{equation}
  P_{z0} = \sum_{\sigma}e\!\!\int_{\Omega_{\text{cell}}}\!\!\! d\text{\bf r}\,z \left[ 
  \psi^*_{\Gamma\sigma}(\text{\bf r})\,\psi_{\Lambda\sigma}(\text{\bf r})\;+\;\text{c.c.}
  \right], 
\end{equation}
the frame origin being placed at the inversion center---the midpoint between the two Mo atoms of the 2$H$ cell. As the charge displacement that gives rise to the dipole is essentially localized on Mo atoms (Fig.~4C), we expect $\left|P_{z0}\right|$ to be well defined. We obtain $P_{z0}/e = $ 15.1 Bohr at $P=34$ GPa.

\end{methods}

\section*{References}

%% Put the bibliography here, most people will use BiBTeX in
%% which case the environment below should be replaced with
%% the \bibliography{} command.

%\bibliography{refs_new,biblio,ei}

\begin{thebibliography}{10}
\expandafter\ifx\csname url\endcsname\relax
  \def\url#1{\texttt{#1}}\fi
\expandafter\ifx\csname urlprefix\endcsname\relax\def\urlprefix{URL }\fi
\providecommand{\bibinfo}[2]{#2}
\providecommand{\eprint}[2][]{\url{#2}}

\bibitem{Keldysh1964}
\bibinfo{author}{Keldysh, L.~V.} \& \bibinfo{author}{Kopaev, Y.~V.}
\newblock \bibinfo{title}{Possible instability of the semimetallic state
  against {C}oulomb interaction}.
\newblock \emph{\bibinfo{journal}{Fiz. Tverd. Tela}}
  \textbf{\bibinfo{volume}{6}}, \bibinfo{pages}{2791} (\bibinfo{year}{1964}).
\newblock \bibinfo{note}{[Sov. Phys. Sol. State {\bf 6,} 2219 (1965)]}.

\bibitem{Cloizeaux1965}
\bibinfo{author}{des Cloizeaux, J.}
\newblock \bibinfo{title}{Excitonic instability and crystallographic anomalies
  in semiconductors}.
\newblock \emph{\bibinfo{journal}{J. Phys. Chem. Solids}}
  \textbf{\bibinfo{volume}{26}}, \bibinfo{pages}{259} (\bibinfo{year}{1965}).

\bibitem{Kohn1967}
\bibinfo{author}{J{\`e}rome, D.}, \bibinfo{author}{Rice, T.~M.} \&
  \bibinfo{author}{Kohn, W.}
\newblock \bibinfo{title}{Excitonic insulator}.
\newblock \emph{\bibinfo{journal}{Phys. Rev.}} \textbf{\bibinfo{volume}{158}},
  \bibinfo{pages}{462} (\bibinfo{year}{1967}).

\bibitem{Kohn1967b}
\bibinfo{author}{Kohn, W.}
\newblock \bibinfo{title}{Metals and insulators}.
\newblock In \bibinfo{editor}{de~Witt, C.} \& \bibinfo{editor}{Balian, R.}
  (eds.) \emph{\bibinfo{booktitle}{Many-body physics}},
  \bibinfo{pages}{351--411} (\bibinfo{publisher}{Gordon and Breach},
  \bibinfo{address}{New York}, \bibinfo{year}{1967}).

\bibitem{BCS1957}
\bibinfo{author}{Bardeen, J.}, \bibinfo{author}{Cooper, L.~N.} \&
  \bibinfo{author}{Schrieffer, J.~R.}
\newblock \bibinfo{title}{Theory of superconductivity}.
\newblock \emph{\bibinfo{journal}{Phys. Rev.}} \textbf{\bibinfo{volume}{108}},
  \bibinfo{pages}{1175--1204} (\bibinfo{year}{1957}).

\bibitem{Halperin1968}
\bibinfo{author}{Halperin, B.~I.} \& \bibinfo{author}{Rice, T.~M.}
\newblock \bibinfo{title}{The excitonic state at the semiconductor-semimetal
  transition}.
\newblock \emph{\bibinfo{journal}{Solid State Phys.}}
  \textbf{\bibinfo{volume}{21}}, \bibinfo{pages}{115} (\bibinfo{year}{1968}).

\bibitem{Guseinov1973}
\bibinfo{author}{Guse\u{\i}nov, R.~R.} \& \bibinfo{author}{Keldysh, L.~V.}
\newblock \bibinfo{title}{Nature of the phase transition under the condition of
  an {``excitonic''} instability in the electronic spectrum of a crystal}.
\newblock \emph{\bibinfo{journal}{Zh. Eksp. i Teor. Fiz.}}
  \textbf{\bibinfo{volume}{63}}, \bibinfo{pages}{2255} (\bibinfo{year}{1972}).
\newblock \bibinfo{note}{[Sov. Phys.--JETP {\bf 36,} 1193 (1973)]}.

\bibitem{Portengen1996b}
\bibinfo{author}{Portengen, T.}, \bibinfo{author}{{\"O}streich, T.} \&
  \bibinfo{author}{Sham, L.~J.}
\newblock \bibinfo{title}{Theory of electronic ferroelectricity}.
\newblock \emph{\bibinfo{journal}{Phys. Rev. B}} \textbf{\bibinfo{volume}{54}},
  \bibinfo{pages}{17452} (\bibinfo{year}{1996}).

\bibitem{Stringari2003}
\bibinfo{author}{Pitaevskii, L.} \& \bibinfo{author}{Stringari, S.}
\newblock \emph{\bibinfo{title}{Bose-Einstein condensation}}
  (\bibinfo{publisher}{Oxford University Press}, \bibinfo{address}{Oxford},
  \bibinfo{year}{2003}).

\bibitem{Eisenstein2004}
\bibinfo{author}{Eisenstein, J.~P.} \& \bibinfo{author}{MacDonald, A.~H.}
\newblock \bibinfo{title}{Bose-{E}instein condensation of excitons in bilayer
  electron systems}.
\newblock \emph{\bibinfo{journal}{Nature}} \textbf{\bibinfo{volume}{432}},
  \bibinfo{pages}{691--694} (\bibinfo{year}{2004}).

\bibitem{Littlewood2008}
\bibinfo{author}{Littlewood, P.~B.}
\newblock \bibinfo{title}{Exciton coherence}.
\newblock In \bibinfo{editor}{Ivanov, A.~L.} \& \bibinfo{editor}{Tikhodeev,
  S.~G.} (eds.) \emph{\bibinfo{booktitle}{Problems of condensed matter
  physics}}, vol. \bibinfo{volume}{139} of \emph{\bibinfo{series}{International
  Series of Monographs on Physics}}, chap.~\bibinfo{chapter}{11},
  \bibinfo{pages}{163--181} (\bibinfo{publisher}{Oxford University Press},
  \bibinfo{address}{Oxford, UK}, \bibinfo{year}{2008}).

\bibitem{Rontani2013}
\bibinfo{author}{Rontani, M.} \& \bibinfo{author}{Sham, L.~J.}
\newblock \bibinfo{title}{Coherent exciton transport in semiconductors}.
\newblock In \bibinfo{editor}{Bennemann, K.~H.} \& \bibinfo{editor}{Ketterson,
  J.~B.} (eds.) \emph{\bibinfo{booktitle}{Novel Superfluids Volume 2}}, vol.
  \bibinfo{volume}{157} of \emph{\bibinfo{series}{International Series of
  Monographs on Physics}}, chap.~\bibinfo{chapter}{19},
  \bibinfo{pages}{423--474} (\bibinfo{publisher}{Oxford University Press},
  \bibinfo{address}{Oxford, UK}, \bibinfo{year}{2014}).

\bibitem{Volkov1973}
\bibinfo{author}{Volkov, V.~A.} \& \bibinfo{author}{Kopaev, Y.~V.}
\newblock \bibinfo{title}{Theory of phase transitions in semiconductors of the
  {A}$_4${B}$_6$ group}.
\newblock \emph{\bibinfo{journal}{Zh. Eksp. i Teor. Fiz.}}
  \textbf{\bibinfo{volume}{64}}, \bibinfo{pages}{2184--2915}
  (\bibinfo{year}{1973}).
\newblock \bibinfo{note}{[Sov. Phys.--JETP {\bf 37,} 1103-1108 (1974)]}.

\bibitem{Varsano2020}
\bibinfo{author}{Varsano, D.}, \bibinfo{author}{Palummo, M.},
  \bibinfo{author}{Molinari, E.} \& \bibinfo{author}{Rontani, M.}
\newblock \bibinfo{title}{A monolayer transition-metal dichalcogenide as a
  topological excitonic insulator}.
\newblock \emph{\bibinfo{journal}{Nature Nanotech.}}
  \textbf{\bibinfo{volume}{15}}, \bibinfo{pages}{367--372}
  (\bibinfo{year}{2020}).
\newblock \urlprefix\url{https://www.nature.com/articles/s41565-020-0650-4}.

\bibitem{Nandi2012}
\bibinfo{author}{Nandi, A.}, \bibinfo{author}{Finck, A. D.~K.},
  \bibinfo{author}{Eisenstein, J.~P.}, \bibinfo{author}{Pfeiffer, L.~N.} \&
  \bibinfo{author}{West, K.~W.}
\newblock \bibinfo{title}{Exciton condensation and perfect {C}oulomb drag}.
\newblock \emph{\bibinfo{journal}{Nature}} \textbf{\bibinfo{volume}{488}},
  \bibinfo{pages}{481} (\bibinfo{year}{2012}).

\bibitem{Butov2002a}
\bibinfo{author}{Butov, L.~V.}, \bibinfo{author}{Lai, C.~W.},
  \bibinfo{author}{Ivanov, A.~L.}, \bibinfo{author}{Gossard, A.~C.} \&
  \bibinfo{author}{Chemla, D.~S.}
\newblock \bibinfo{title}{Towards {B}ose–{E}instein condensation of excitons
  in potential traps}.
\newblock \emph{\bibinfo{journal}{Nature}} \textbf{\bibinfo{volume}{417}},
  \bibinfo{pages}{47--52} (\bibinfo{year}{2002}).

\bibitem{High2012}
\bibinfo{author}{High, A.~A.} \emph{et~al.}
\newblock \bibinfo{title}{Spontaneous coherence in a cold exciton gas}.
\newblock \emph{\bibinfo{journal}{Nature}} \textbf{\bibinfo{volume}{483}},
  \bibinfo{pages}{584--588} (\bibinfo{year}{2012}).

\bibitem{Anankine2017}
\bibinfo{author}{Anankine, R.} \emph{et~al.}
\newblock \bibinfo{title}{Quantized vortices and four-component superfluidity
  of semiconductor excitons}.
\newblock \emph{\bibinfo{journal}{Phys. Rev. Lett.}}
  \textbf{\bibinfo{volume}{118}}, \bibinfo{pages}{127402}
  (\bibinfo{year}{2017}).
\newblock
  \urlprefix\url{https://link.aps.org/doi/10.1103/PhysRevLett.118.127402}.

\bibitem{Rohwer2011}
\bibinfo{author}{Rohwer, T.} \emph{et~al.}
\newblock \bibinfo{title}{Collapse of long-range charge order tracked by
  time-resolved photoemission at high momenta}.
\newblock \emph{\bibinfo{journal}{Nature}} \textbf{\bibinfo{volume}{471}},
  \bibinfo{pages}{490--494} (\bibinfo{year}{2011}).

\bibitem{Kogar-Abbamonte_2017}
\bibinfo{author}{Kogar, A.} \emph{et~al.}
\newblock \bibinfo{title}{Signatures of exciton condensation in a transition
  metal dichalcogenide}.
\newblock \emph{\bibinfo{journal}{Science}} \textbf{\bibinfo{volume}{358}},
  \bibinfo{pages}{1314--1317} (\bibinfo{year}{2017}).

\bibitem{Kono2017}
\bibinfo{author}{Lu, Y.~F.} \emph{et~al.}
\newblock \bibinfo{title}{{Zero-gap semiconductor to excitonic insulator
  transition in Ta$_2$NiSe$_5$}}.
\newblock \emph{\bibinfo{journal}{{Nature Commun.}}}
  \textbf{\bibinfo{volume}{{8}}}, \bibinfo{pages}{{14408}}
  (\bibinfo{year}{{2017}}).

\bibitem{Kaiser2018}
\bibinfo{author}{Werdehausen, D.} \emph{et~al.}
\newblock \bibinfo{title}{Coherent order parameter oscillations in the ground
  state of the excitonic insulator {T}a$_2${N}i{S}e$_5$}.
\newblock \emph{\bibinfo{journal}{Science Adv.}} \textbf{\bibinfo{volume}{4}},
  \bibinfo{pages}{eaap8652} (\bibinfo{year}{2018}).

\bibitem{DiSalvo1976}
\bibinfo{author}{Salvo, F. J.~D.}, \bibinfo{author}{Moncton, D.~E.} \&
  \bibinfo{author}{Waszczak, J.~V.}
\newblock \bibinfo{title}{Electronic properties and superlattice formation in
  the semimetal {T}i{S}e$_2$}.
\newblock \emph{\bibinfo{journal}{Phys. Rev. B}} \textbf{\bibinfo{volume}{14}},
  \bibinfo{pages}{4321} (\bibinfo{year}{1976}).

\bibitem{Hedayat2019}
\bibinfo{author}{Hedayat, H.} \emph{et~al.}
\newblock \bibinfo{title}{Excitonic and lattice contributions to the charge
  density wave in {$1T$}-{T}i{S}e$_{2}$ revealed by a phonon bottleneck}.
\newblock \emph{\bibinfo{journal}{Phys. Rev. Research}}
  \textbf{\bibinfo{volume}{1}}, \bibinfo{pages}{023029} (\bibinfo{year}{2019}).
\newblock
  \urlprefix\url{https://link.aps.org/doi/10.1103/PhysRevResearch.1.023029}.

\bibitem{Zhou2019}
\bibinfo{author}{Zhou, J.~S.} \emph{et~al.}
\newblock \bibinfo{title}{Anharmonic melting of the charge density wave in
  single-layer {T}i{S}e$_2$} (\bibinfo{year}{2019}).
\newblock \urlprefix\url{https://arxiv.org/abs/1910.12709}.
\newblock \bibinfo{note}{{a}rXiv:1910.12709}.

\bibitem{Nakano2018}
\bibinfo{author}{Nakano, A.} \emph{et~al.}
\newblock \bibinfo{title}{Antiferroelectric distortion with anomalous phonon
  softening in the excitonic insulator {T}a$_2${N}i{S}e$_5$}.
\newblock \emph{\bibinfo{journal}{Phys. Rev. B}} \textbf{\bibinfo{volume}{98}},
  \bibinfo{pages}{045139} (\bibinfo{year}{2018}).
\newblock \urlprefix\url{https://link.aps.org/doi/10.1103/PhysRevB.98.045139}.

\bibitem{Yan2019}
\bibinfo{author}{Yan, J.} \emph{et~al.}
\newblock \bibinfo{title}{Strong electron-phonon coupling in the excitonic
  insulator {T}a$_2${N}i{S}e$_5$}.
\newblock \emph{\bibinfo{journal}{Inorganic Chemistry}}
  \textbf{\bibinfo{volume}{58}}, \bibinfo{pages}{9036--9042}
  (\bibinfo{year}{2019}).
\newblock \urlprefix\url{https://doi.org/10.1021/acs.inorgchem.9b00432}.
\newblock \bibinfo{note}{PMID: 31246443},
  \eprint{https://doi.org/10.1021/acs.inorgchem.9b00432}.

\bibitem{Kohn1970}
\bibinfo{author}{Kohn, W.} \& \bibinfo{author}{Sherrington, D.}
\newblock \bibinfo{title}{Two kinds of bosons and {B}ose condensates}.
\newblock \emph{\bibinfo{journal}{Rev. Mod. Phys.}}
  \textbf{\bibinfo{volume}{42}}, \bibinfo{pages}{1} (\bibinfo{year}{1970}).

\bibitem{tosatti}
\bibinfo{author}{Hromadov\'a, L.},
  \bibinfo{author}{Marto\ifmmode~\check{n}\else \v{n}\fi{}\'ak, R.} \&
  \bibinfo{author}{Tosatti, E.}
\newblock \bibinfo{title}{Structure change, layer sliding, and metallization in
  high-pressure {M}o{S}${}_{2}$}.
\newblock \emph{\bibinfo{journal}{Phys. Rev. B}} \textbf{\bibinfo{volume}{87}},
  \bibinfo{pages}{144105} (\bibinfo{year}{2013}).
\newblock \urlprefix\url{https://link.aps.org/doi/10.1103/PhysRevB.87.144105}.

\bibitem{Chi2014}
\bibinfo{author}{Chi, Z.-H.} \emph{et~al.}
\newblock \bibinfo{title}{Pressure-induced metallization of molybdenum
  disulfide}.
\newblock \emph{\bibinfo{journal}{Phys. Rev. Lett.}}
  \textbf{\bibinfo{volume}{113}}, \bibinfo{pages}{036802}
  (\bibinfo{year}{2014}).
\newblock
  \urlprefix\url{https://link.aps.org/doi/10.1103/PhysRevLett.113.036802}.

\bibitem{Nayak2014}
\bibinfo{author}{Nayak, A.~P.} \emph{et~al.}
\newblock \bibinfo{title}{Pressure-induced semiconducting to metallic
  transition in multilayered molybdenum disulphide}.
\newblock \emph{\bibinfo{journal}{Nature Commun.}}
  \textbf{\bibinfo{volume}{5}}, \bibinfo{pages}{3731} (\bibinfo{year}{2014}).

\bibitem{Chi2018}
\bibinfo{author}{Chi, Z.} \emph{et~al.}
\newblock \bibinfo{title}{Superconductivity in pristine
  $2{H}_{a}$-{M}o{S}$_{2}$ at ultrahigh pressure}.
\newblock \emph{\bibinfo{journal}{Phys. Rev. Lett.}}
  \textbf{\bibinfo{volume}{120}}, \bibinfo{pages}{037002}
  (\bibinfo{year}{2018}).
\newblock
  \urlprefix\url{https://link.aps.org/doi/10.1103/PhysRevLett.120.037002}.

\bibitem{Onida-Reining-Rubio_2002}
\bibinfo{author}{Onida, G.}, \bibinfo{author}{Reining, L.} \&
  \bibinfo{author}{Rubio, A.}
\newblock \bibinfo{title}{{Electronic excitations: density-functional versus
  many-body Green's-function approaches}}.
\newblock \emph{\bibinfo{journal}{Rev. Mod. Phys.}}
  \textbf{\bibinfo{volume}{74}}, \bibinfo{pages}{601--659}
  (\bibinfo{year}{2002}).

\bibitem{Varsano-Rontani2017}
\bibinfo{author}{Varsano, D.} \emph{et~al.}
\newblock \bibinfo{title}{{Carbon nanotubes as excitonic insulators}}.
\newblock \emph{\bibinfo{journal}{Nature Commun.}}
  \textbf{\bibinfo{volume}{8}}, \bibinfo{pages}{1461} (\bibinfo{year}{2017}).

\bibitem{Cao2018}
\bibinfo{author}{Cao, Z.-Y.}, \bibinfo{author}{Hu, J.-W.},
  \bibinfo{author}{Goncharov, A.~F.} \& \bibinfo{author}{Chen, X.-J.}
\newblock \bibinfo{title}{Nontrivial metallic state of {M}o{S}$_2$}.
\newblock \emph{\bibinfo{journal}{Phys. Rev. B}} \textbf{\bibinfo{volume}{97}},
  \bibinfo{pages}{214519} (\bibinfo{year}{2018}).
\newblock \urlprefix\url{https://link.aps.org/doi/10.1103/PhysRevB.97.214519}.

\bibitem{Ge2013}
\bibinfo{author}{Ge, Y.} \& \bibinfo{author}{Liu, A.~Y.}
\newblock \bibinfo{title}{Phonon-mediated superconductivity in electron-doped
  single-layer {M}o{S}$_2$: A first-principles prediction}.
\newblock \emph{\bibinfo{journal}{Phys. Rev. B}} \textbf{\bibinfo{volume}{87}},
  \bibinfo{pages}{241408} (\bibinfo{year}{2013}).
\newblock \urlprefix\url{https://link.aps.org/doi/10.1103/PhysRevB.87.241408}.

\bibitem{Roldan2013}
\bibinfo{author}{Rold\'an, R.}, \bibinfo{author}{Cappelluti, E.} \&
  \bibinfo{author}{Guinea, F.}
\newblock \bibinfo{title}{Interactions and superconductivity in heavily doped
  {M}o{S}$_2$}.
\newblock \emph{\bibinfo{journal}{Phys. Rev. B}} \textbf{\bibinfo{volume}{88}},
  \bibinfo{pages}{054515} (\bibinfo{year}{2013}).
\newblock \urlprefix\url{https://link.aps.org/doi/10.1103/PhysRevB.88.054515}.

\bibitem{Rosner2014}
\bibinfo{author}{R\"osner, M.}, \bibinfo{author}{Haas, S.} \&
  \bibinfo{author}{Wehling, T.~O.}
\newblock \bibinfo{title}{Phase diagram of electron-doped dichalcogenides}.
\newblock \emph{\bibinfo{journal}{Phys. Rev. B}} \textbf{\bibinfo{volume}{90}},
  \bibinfo{pages}{245105} (\bibinfo{year}{2014}).
\newblock \urlprefix\url{https://link.aps.org/doi/10.1103/PhysRevB.90.245105}.

\bibitem{aksoy}
\bibinfo{author}{Aksoy, R.} \emph{et~al.}
\newblock \bibinfo{title}{X-ray diffraction study of molybdenum disulfide to
  38.8 {GP}a}.
\newblock \emph{\bibinfo{journal}{J. Phys. Chem. Solids}}
  \textbf{\bibinfo{volume}{67}}, \bibinfo{pages}{1914--1917}
  (\bibinfo{year}{2006}).

\bibitem{Bandaru2014}
\bibinfo{author}{Bandaru, N.} \emph{et~al.}
\newblock \bibinfo{title}{Effect of pressure and temperature on structural
  stability of {M}o{S}$_2$}.
\newblock \emph{\bibinfo{journal}{J. Phys. Chem. C}}
  \textbf{\bibinfo{volume}{118}}, \bibinfo{pages}{3230--3235}
  (\bibinfo{year}{2014}).

\bibitem{Zhuang2017}
\bibinfo{author}{Zhuang, Y.} \emph{et~al.}
\newblock \bibinfo{title}{Pressure-induced permanent metallization with
  reversible structural transition in molybdenum disulfide}.
\newblock \emph{\bibinfo{journal}{Applied Physics Letters}}
  \textbf{\bibinfo{volume}{110}}, \bibinfo{pages}{122103}
  (\bibinfo{year}{2017}).
\newblock \urlprefix\url{https://doi.org/10.1063/1.4979143}.
\newblock \eprint{https://doi.org/10.1063/1.4979143}.

\bibitem{ordejon}
\bibinfo{author}{Brotons-Gisbert, M.} \emph{et~al.}
\newblock \bibinfo{title}{Optical and electronic properties of
  $2${H}-{M}o{S}$_2$ under pressure: Revealing the spin-polarized nature of
  bulk electronic bands}.
\newblock \emph{\bibinfo{journal}{Phys. Rev. Materials}}
  \textbf{\bibinfo{volume}{2}}, \bibinfo{pages}{054602} (\bibinfo{year}{2018}).
\newblock
  \urlprefix\url{https://link.aps.org/doi/10.1103/PhysRevMaterials.2.054602}.

\bibitem{Goncharov2020}
\bibinfo{author}{Goncharov, A.~F.} \emph{et~al.}
\newblock \bibinfo{title}{Structure and stability of $2{H}_{a}$-{M}o{S}$_2$ at
  high pressure and low temperatures}.
\newblock \emph{\bibinfo{journal}{Phys. Rev. B}}
  \textbf{\bibinfo{volume}{102}}, \bibinfo{pages}{064105}
  (\bibinfo{year}{2020}).
\newblock \urlprefix\url{https://link.aps.org/doi/10.1103/PhysRevB.102.064105}.

\bibitem{guo}
\bibinfo{author}{Guo, H.}, \bibinfo{author}{Yang, T.}, \bibinfo{author}{Tao,
  P.}, \bibinfo{author}{Wang, Y.} \& \bibinfo{author}{Zhang, Z.}
\newblock \bibinfo{title}{High pressure effect on structure, electronic
  structure, and thermoelectric properties of {M}o{S}$_2$}.
\newblock \emph{\bibinfo{journal}{J. Appl. Phys.}}
  \textbf{\bibinfo{volume}{113}}, \bibinfo{pages}{013709}
  (\bibinfo{year}{2013}).

\bibitem{Monney2009}
\bibinfo{author}{Monney, C.} \emph{et~al.}
\newblock \bibinfo{title}{Spontaneous exciton condensation in
  1{T}-{T}i{S}e$_2$: {BCS}-like approach}.
\newblock \emph{\bibinfo{journal}{Phys. Rev. B}} \textbf{\bibinfo{volume}{79}},
  \bibinfo{pages}{045116} (\bibinfo{year}{2009}).
\newblock \urlprefix\url{https://link.aps.org/doi/10.1103/PhysRevB.79.045116}.

\bibitem{Mattheiss1973}
\bibinfo{author}{Mattheiss, L.~F.}
\newblock \bibinfo{title}{Band structures of transition-metal-dichalcogenide
  layer compounds}.
\newblock \emph{\bibinfo{journal}{Phys. Rev. B}} \textbf{\bibinfo{volume}{8}},
  \bibinfo{pages}{3719--3740} (\bibinfo{year}{1973}).

\bibitem{Kozlov1965}
\bibinfo{author}{Kozlov, A.~N.} \& \bibinfo{author}{Maksimov, L.~A.}
\newblock \bibinfo{title}{The metal-dielectric divalent crystal phase
  transition}.
\newblock \emph{\bibinfo{journal}{Zh. Eksp. i Teor. Fiz.}}
  \textbf{\bibinfo{volume}{48}}, \bibinfo{pages}{1184--1193}
  (\bibinfo{year}{1965}).
\newblock \bibinfo{note}{[Sov. Phys.--JETP {\bf 21,} 790-795 (1965)]}.

\bibitem{Knox1963}
\bibinfo{author}{Knox, R.~S.}
\newblock \emph{\bibinfo{title}{Theory of excitons}}, vol.
  \bibinfo{volume}{Supplement 5} of \emph{\bibinfo{series}{Solid State
  Physics}} (\bibinfo{publisher}{Academic Press}, \bibinfo{address}{New York},
  \bibinfo{year}{1963}).

\bibitem{Gruner2018}
\bibinfo{author}{Gr{\"u}ner, G.}
\newblock \emph{\bibinfo{title}{Density waves in solids}}
  (\bibinfo{publisher}{CRC Press}, \bibinfo{address}{Boca Raton},
  \bibinfo{year}{2018}).

\bibitem{Kozlov1965b}
\bibinfo{author}{Kozlov, A.~N.} \& \bibinfo{author}{Maksimov, L.~A.}
\newblock \bibinfo{title}{Collective excitations in semimetals}.
\newblock \emph{\bibinfo{journal}{Zh. Eksp. i Teor. Fiz.}}
  \textbf{\bibinfo{volume}{49}}, \bibinfo{pages}{1284--1292}
  (\bibinfo{year}{1965}).
\newblock \bibinfo{note}{[Sov. Phys.--JETP {\bf 22,} 889-893 (1966)]}.

\bibitem{Giannozzi2009}
\bibinfo{author}{Giannozzi, P.} \emph{et~al.}
\newblock \bibinfo{title}{Quantum {ESPRESSO}: a modular and open-source
  software project for quantum simulations of materials}.
\newblock \emph{\bibinfo{journal}{J. Phys.: Condens. Matter}}
  \textbf{\bibinfo{volume}{21}}, \bibinfo{pages}{395502}
  (\bibinfo{year}{2009}).

\bibitem{Giannozzi2017}
\bibinfo{author}{Giannozzi, P.} \emph{et~al.}
\newblock \bibinfo{title}{Advanced capabilities for materials modelling with
  quantum espresso}.
\newblock \emph{\bibinfo{journal}{J. Phys.: Condens. Matter}}
  \textbf{\bibinfo{volume}{29}}, \bibinfo{pages}{465901}
  (\bibinfo{year}{2017}).

\bibitem{pbe96}
\bibinfo{author}{Perdew, J.~P.}, \bibinfo{author}{Burke, K.} \&
  \bibinfo{author}{Ernzerhof, M.}
\newblock \bibinfo{title}{Generalized gradient approximation made simple}.
\newblock \emph{\bibinfo{journal}{Physical Review Letters}}
  \textbf{\bibinfo{volume}{77}}, \bibinfo{pages}{3865} (\bibinfo{year}{1996}).

\bibitem{hamann2013optimized}
\bibinfo{author}{Hamann, D.}
\newblock \bibinfo{title}{Optimized norm-conserving {V}anderbilt
  pseudopotentials}.
\newblock \emph{\bibinfo{journal}{Physical Review B}}
  \textbf{\bibinfo{volume}{88}}, \bibinfo{pages}{085117}
  (\bibinfo{year}{2013}).

\bibitem{Baroni2001}
\bibinfo{author}{Baroni, S.}, \bibinfo{author}{de~Gironcoli, S.},
  \bibinfo{author}{Dal~Corso, A.} \& \bibinfo{author}{Giannozzi, P.}
\newblock \bibinfo{title}{Phonons and related crystal properties from
  density-functional perturbation theory}.
\newblock \emph{\bibinfo{journal}{Rev. Mod. Phys.}}
  \textbf{\bibinfo{volume}{73}}, \bibinfo{pages}{515--562}
  (\bibinfo{year}{2001}).
\newblock \urlprefix\url{https://link.aps.org/doi/10.1103/RevModPhys.73.515}.

\bibitem{hybertsen1986}
\bibinfo{author}{Hybertsen, M.~S.} \& \bibinfo{author}{Louie, S.~G.}
\newblock \bibinfo{title}{Electron correlation in semiconductors and
  insulators: Band gaps and quasiparticle energies}.
\newblock \emph{\bibinfo{journal}{Phys. Rev. B}} \textbf{\bibinfo{volume}{34}},
  \bibinfo{pages}{5390} (\bibinfo{year}{1986}).

\bibitem{strinati1988g}
\bibinfo{author}{Strinati, G.}
\newblock \bibinfo{title}{Application of the {G}reen’s functions method to
  the study of the optical properties of semiconductors}.
\newblock \emph{\bibinfo{journal}{Riv. Nuovo Cimento}}
  \textbf{\bibinfo{volume}{11}}, \bibinfo{pages}{1} (\bibinfo{year}{1988}).

\bibitem{Marini2009}
\bibinfo{author}{Marini, A.}, \bibinfo{author}{Hogan, C.},
  \bibinfo{author}{Gr{\"u}ning, M.} \& \bibinfo{author}{Varsano, D.}
\newblock \bibinfo{title}{Yambo: {A}n ab initio tool for excited state
  calculations}.
\newblock \emph{\bibinfo{journal}{Comput. Phys. Commun.}}
  \textbf{\bibinfo{volume}{180}}, \bibinfo{pages}{1392--1403}
  (\bibinfo{year}{2009}).

\bibitem{yambo}
\bibinfo{author}{Sangalli, D.} \emph{et~al.}
\newblock \bibinfo{title}{Many-body perturbation theory calculations using the
  {Y}ambo code}.
\newblock \emph{\bibinfo{journal}{Journal of Physics: Condensed Matter}}
  \textbf{\bibinfo{volume}{31}}, \bibinfo{pages}{325902}
  (\bibinfo{year}{2019}).

\bibitem{godby}
\bibinfo{author}{Godby, R.~W.} \& \bibinfo{author}{Needs, R.~J.}
\newblock \bibinfo{title}{Metal-insulator transition in {K}ohn-{S}ham theory
  and quasiparticle theory}.
\newblock \emph{\bibinfo{journal}{Phys. Rev. Lett.}}
  \textbf{\bibinfo{volume}{62}}, \bibinfo{pages}{1169--1172}
  (\bibinfo{year}{1989}).
\newblock \urlprefix\url{https://link.aps.org/doi/10.1103/PhysRevLett.62.1169}.

\bibitem{bruneval2008accurate}
\bibinfo{author}{Bruneval, F.} \& \bibinfo{author}{Gonze, X.}
\newblock \bibinfo{title}{Accurate {GW} self-energies in a plane-wave basis
  using only a few empty states: Towards large systems}.
\newblock \emph{\bibinfo{journal}{Physical Review B}}
  \textbf{\bibinfo{volume}{78}}, \bibinfo{pages}{085125}
  (\bibinfo{year}{2008}).

\bibitem{gatti2013}
\bibinfo{author}{Gatti, M.} \& \bibinfo{author}{Sottile, F.}
\newblock \bibinfo{title}{Exciton dispersion from first principles}.
\newblock \emph{\bibinfo{journal}{Phys. Rev. B}} \textbf{\bibinfo{volume}{88}},
  \bibinfo{pages}{155113} (\bibinfo{year}{2013}).
\newblock \urlprefix\url{https://link.aps.org/doi/10.1103/PhysRevB.88.155113}.

\bibitem{Shirley2000}
\bibinfo{author}{Soininen, J.~A.} \& \bibinfo{author}{Shirley, E.~L.}
\newblock \bibinfo{title}{Effects of electron-hole interaction on the dynamic
  structure factor: Application to nonresonant inelastic x-ray scattering}.
\newblock \emph{\bibinfo{journal}{Phys. Rev. B}} \textbf{\bibinfo{volume}{61}},
  \bibinfo{pages}{16423--16429} (\bibinfo{year}{2000}).
\newblock \urlprefix\url{https://link.aps.org/doi/10.1103/PhysRevB.61.16423}.

\bibitem{Pikus1974}
\bibinfo{author}{Bir, G.~L.} \& \bibinfo{author}{Pikus, G.~E.}
\newblock \emph{\bibinfo{title}{Symmetry and Strain-Induced Effects in
  Semiconductors}} (\bibinfo{publisher}{Wiley}, \bibinfo{address}{New York},
  \bibinfo{year}{1974}).

\bibitem{Pedersen2016}
\bibinfo{author}{Pedersen, T.~G.}, \bibinfo{author}{Latini, S.},
  \bibinfo{author}{Thygesen, K.~S.}, \bibinfo{author}{Mera, H.} \&
  \bibinfo{author}{Nikoli{\'{c}}, B.~K.}
\newblock \bibinfo{title}{Exciton ionization in multilayer transition-metal
  dichalcogenides}.
\newblock \emph{\bibinfo{journal}{New J. Phys.}} \textbf{\bibinfo{volume}{18}},
  \bibinfo{pages}{073043} (\bibinfo{year}{2016}).
\newblock
  \urlprefix\url{https://iopscience.iop.org/article/10.1088/1367-2630/18/7/073043}.

\bibitem{Zittartz1967}
\bibinfo{author}{Zittartz, J.}
\newblock \bibinfo{title}{Anisotropy effects in the excitonic insulator}.
\newblock \emph{\bibinfo{journal}{Phys. Rev.}} \textbf{\bibinfo{volume}{162}},
  \bibinfo{pages}{752--758} (\bibinfo{year}{1967}).

\end{thebibliography}

%reference.

%% Here is the endmatter stuff: Supplementary Info, etc.
%% Use \item's to separate, default label is "Acknowledgements"

\begin{addendum}
 \item[Supplementary Information] is available in the online
 version of the paper.
 \item D.V.~acknowledges the joint work with Davide Sangalli to implement the finite-momentum Bethe-Salpeter calculation into the Yambo code. This work was supported in part by the MaX European Centre of Excellence: MaX (``MAterials design at the eXascale'', www.max-centre.eu) funded by the European Union H2020-INFRAEDI-2018-1 programme, grant No. 824143. It was also supported by the Italian national program PRIN2017 No.~2017BZPKSZ `Excitonic insulator in two-dimensional long-range interacting systems (EXC-INS)'. The authors acknowledge access to the Marconi supercomputing system based at CINECA, Italy, through PRACE as well as the Italian ISCRA program. 
 \item[Competing Interests] The authors declare that they have no
 competing financial interests.
 \item[Correspondence] Correspondence and requests for materials
 should be addressed to M.R.~(email: massimo.rontani@nano.cnr.it).
\end{addendum}

%%
%% TABLES
%%
%% If there are any tables, put them here.
%%

\end{document}